\documentclass{aa}
\usepackage{graphicx}
\usepackage{txfonts}
\usepackage{natbib}
\usepackage{color}

\bibpunct{(}{)}{;}{a}{}{,}

\begin{document}

\title{Hydrodynamical simulations of the jet\\ in the symbiotic star MWC 560}
\subtitle{II. Simulations beyond density balance}

\titlerunning{Hydrodynamical simulations of the jet in MWC 560. II.}

\author{Matthias Stute\thanks{\emph{Present address:} Jet Propulsion 
Laboratory, Mail Stop 169-506, 4800 Oak Grove
Drive, Pasadena, CA 91109, USA}}
\institute{Landessternwarte Heidelberg, K\"onigstuhl, D-69117 Heidelberg, 
Germany}

\offprints{Matthias Stute, \email{Matthias.Stute@jpl.nasa.gov}}

\date{Received 23 August 2005 / Accepted 19 December 2005}

\abstract
{In the first paper of this series, we presented hydrodynamical simulations 
with radiative cooling of jet models with parameters representative of the 
symbiotic system MWC 560. These were jet simulations of a pulsed, initially 
underdense jet in a high-density ambient medium. They were stopped when the 
jet reached a length of 50 AU. There, however, a transition of the initially 
underdense jet towards an overdense jet should occur, which should result in 
changed kinematics. A few minor differences between the models and the 
observations were thought to be solved by a model with an increased jet 
density during the pulses which was calculated only with purely hydrodynamical 
means in the former paper.}
{Therefore, we describe two hydrodynamical simulations with cooling beyond 
this density balance, one with the same parameters as model i in Paper I 
(now called model i'), which was presented there with and without cooling, and 
the second with higher gas densities in the jet pulses (model iv').}
{Hydrodynamical simulations, with a further approximated cooling treatment 
compared to Paper I, were used to be able to enlarge the computational domain.}
{The transition causes changes in the expansion of the cocoon and
therefore the morphology of the jet, e.g. a larger radial width of the jet 
knots. We investigate the radiation properties of the jets, the 
bremsstrahlung and optical emissivities, integrated emission maps, and 
synthetic absorption line profiles.}
{The conclusion that the high observed velocities in CH Cygni, R 
Aquarii, and MWC 560 favor the models with cooling is unchanged by the 
transition. The observed parallel features in R Aquarii can be produced by the
internal knots or by a variable dense radiative shell of shocked ambient 
medium. The absorption line profiles show that the real parameters in MWC 560 
are closer to model iv' than to model i'.}


\maketitle

\section{Introduction}

The emergence of jets is a very common phenomenon in a variety of 
astrophysical objects, and it occurs in systems of very different size and 
mass scales. Jets are observed in active galactic nuclei (AGN) in which they 
are formed around supermassive black holes, in X-Ray binaries with stellar 
black holes or neutron stars, in young stellar objects (YSO), in supersoft 
X-ray sources and in symbiotic stars. Symbiotic systems consist of a red giant 
undergoing strong mass loss and a white dwarf. More than one hundred symbiotic 
stars are known, but only about ten systems show jet emission. The most famous 
systems are R Aquarii, CH Cygni, and MWC 560. 

R Aquarii, with a distance of about 200 pc, is one of the nearest 
symbiotic stars and a well known jet source. The jet has been extensively 
observed in the optical, at radio wavelengths, and with X-ray observations
\citep[e.g.][]{SoU,PaH,HMKM,HLDF,KPL}. It shows a rich morphology, e.g. 
a series of parallel features in the jet and the counter-jet, extending a few 
hundred AU each. In 1984/85 CH Cygni showed a strong radio outburst, during 
which a double-sided jet with multiple components was ejected \citep{TSM}. 
This event allowed an accurate measurement of the jet velocity near 1500 km 
s$^{-1}$. In HST observations \citep{EBS}, arcs can be detected that also 
could be produced by episodic ejection events. While the first two objects are 
seen at high inclinations, the jet axis in MWC 560 is practically parallel to 
the line of sight. This special orientation provides an opportunity to 
observe the outflowing gas as line absorption in the source spectrum. With 
such observations the radial velocity and the column density of the outflowing 
jet gas close to the source can be investigated in great detail. In particular 
we can probe the acceleration and evolution of individual outflow components 
with spectroscopic monitoring programs, as described in \citet{SKC}.

In \citet[][hereafter Paper I]{SCS}, we presented hydrodynamical 
simulations with and without cooling of jets with parameters that were 
intended to represent those in MWC 560. In a grid of eight simulations we 
investigated the influence of different jet pulse parameters. Due to the 
high computational costs of simulations including cooling, this grid was 
restricted to adiabatic simulations. Only one model simulation was performed 
that included a treatment of radiative cooling. 

The adiabatic jet models showed some quantitative differences for 
the different pulse parameters. In comparison with the model including cooling,
huge differences were present. Those concern, e.g., a significantly higher 
bow-shock velocity and a well-defined periodic shock structure with 
high-density, low-temperature knots separated by hot, low-density beam 
sections in the radiative model. The expansion velocity of CH Cyg was shown to 
agree very closely with the jet model with cooling. The situation in R Aqr was 
less clear. The observationally derived proper motion was more in the range of 
the gas motion of the adiabatic models, although the origin of the observed 
emission has not been confirmed yet. The small jet opening angle of 
15$^{\circ}$ favored a jet with cooling more. We showed that the adiabatic jet 
models are not able to produce low temperature regions, which are, however, 
necessary to explain strong absorptions from low ionization species in the 
spectra of MWC 560 as \ion{Ca}{ii}, but the cooling of the jet gas still did 
not seem efficient enough. This should be achieved by another model with 
higher gas densities in the jet outflow. The basic structure of the observed 
jet absorptions in MWC 560 was reproduced, including the mean velocity, the 
velocity width and temporal evolution of the highest velocity components, 
but not their strengths. Again, another model with higher gas densities 
should solve this.

A drawback of our former simulations was the fact that they were stopped when 
the jet reached a length of 50 AU. This is unsatisfactory from an 
observational point of view, as the observed extents of the jet in R Aquarii 
and CH Cygni are much larger, as well as from the theoretical viewpoint. At 
this point, the density of the environment has decreased down to the density 
in the jet nozzle. A transition of the initially underdense jet towards an 
overdense jet should occur, which should result in changed kinematics. 
Underdense jets decelerate more than overdense jets, because the position of 
the jet head can be basically derived from a balance of the ram pressures of 
jet and external medium and because the relative importance of ram and thermal 
pressure should also change. As the kinematics of the models was one of the 
results of the former models, which were compared with the observations, it 
should be tested whether simulations with larger spatial scales lead to new 
insights into the physics of jets in symbiotic stars.

Therefore, we performed two hydrodynamical simulations with cooling beyond 
this density balance, one with the same parameters as model i in Paper I, 
which was presented both with and without cooling, and the second with higher 
gas densities in the jet pulses. Due to constraints set by the available 
computer resources, we were forced to approximate the former treatment of 
radiative cooling again. 

In Sect. \ref{sec_model}, we briefly describe the changes of our model 
with respect to Paper I. After a validating of the approximated cooling 
treatment in section \ref{sec_valid}, we investigate the global and internal 
structure of the jet (section \ref{sec_struct}) and its consequences on the 
emission (section \ref{sec_emiss}). Then synthetic absorption line profiles 
are calculated in section \ref{sec_abs}. Finally a summary and a discussion 
are given.

\section{The numerical models} \label{sec_model}

\subsection{The computer code}

\begin{figure*}
  \centering
  \includegraphics[width=8cm]{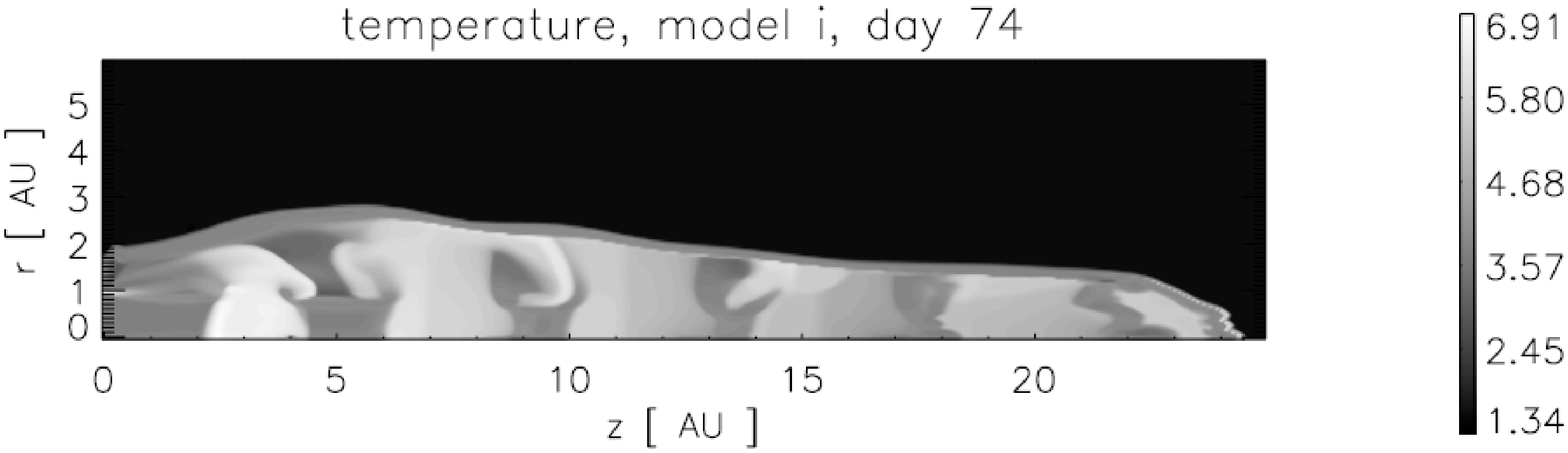}
  \includegraphics[width=8cm]{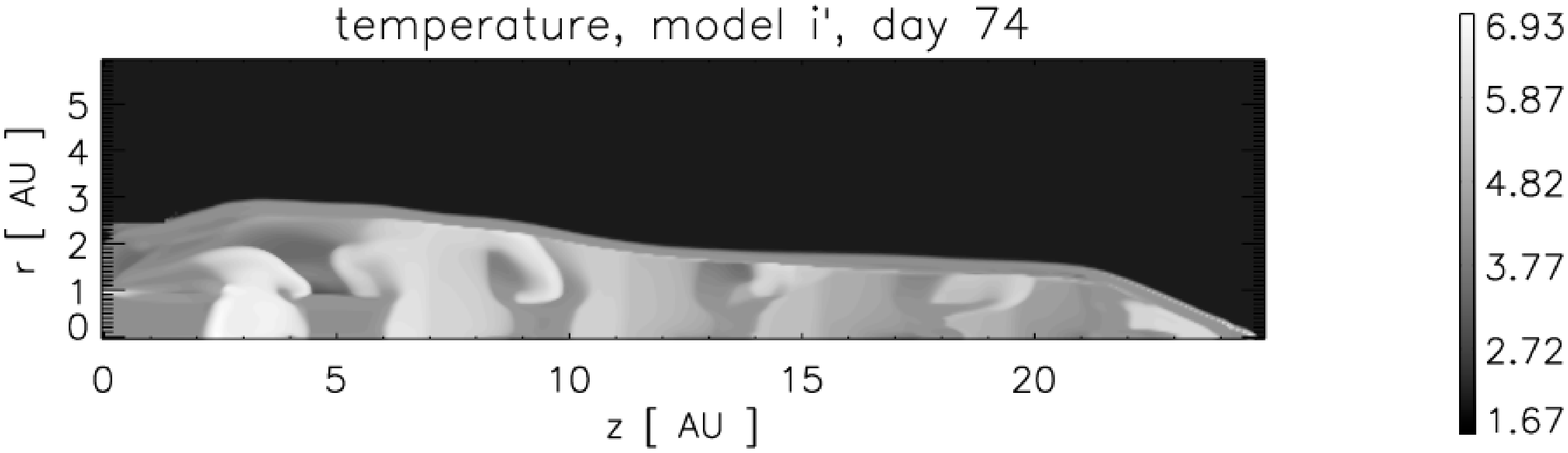}

  \includegraphics[width=8cm]{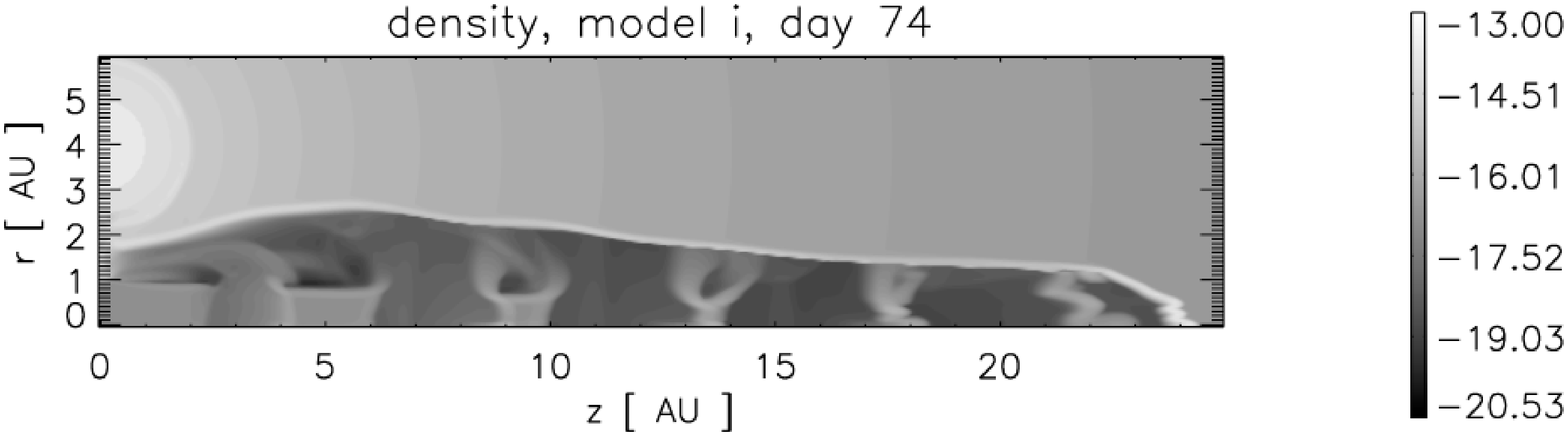}
  \includegraphics[width=8cm]{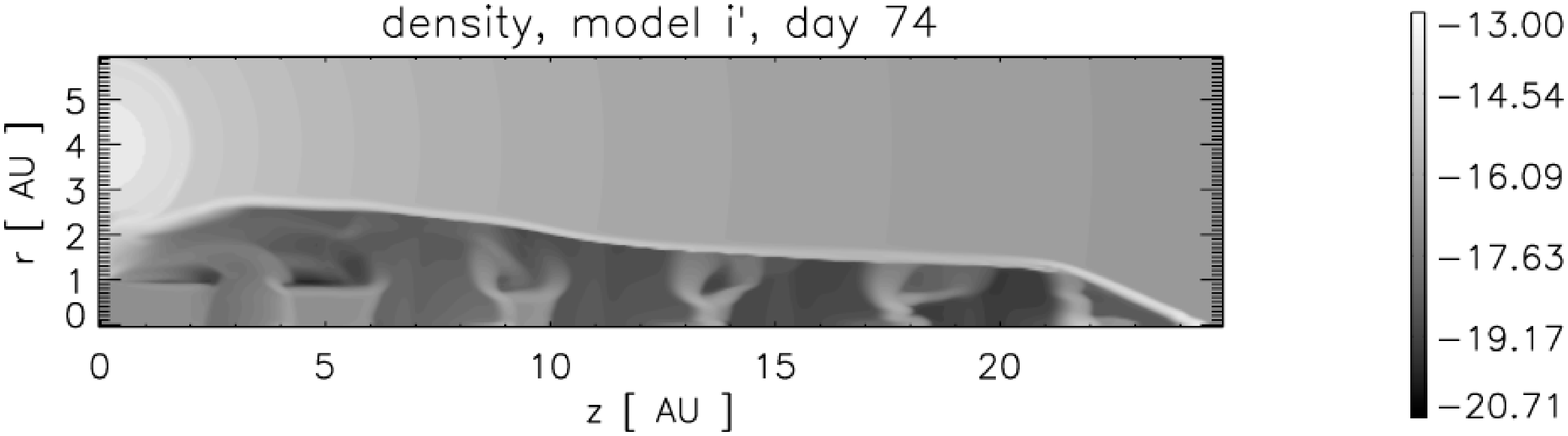}

  \includegraphics[width=8cm]{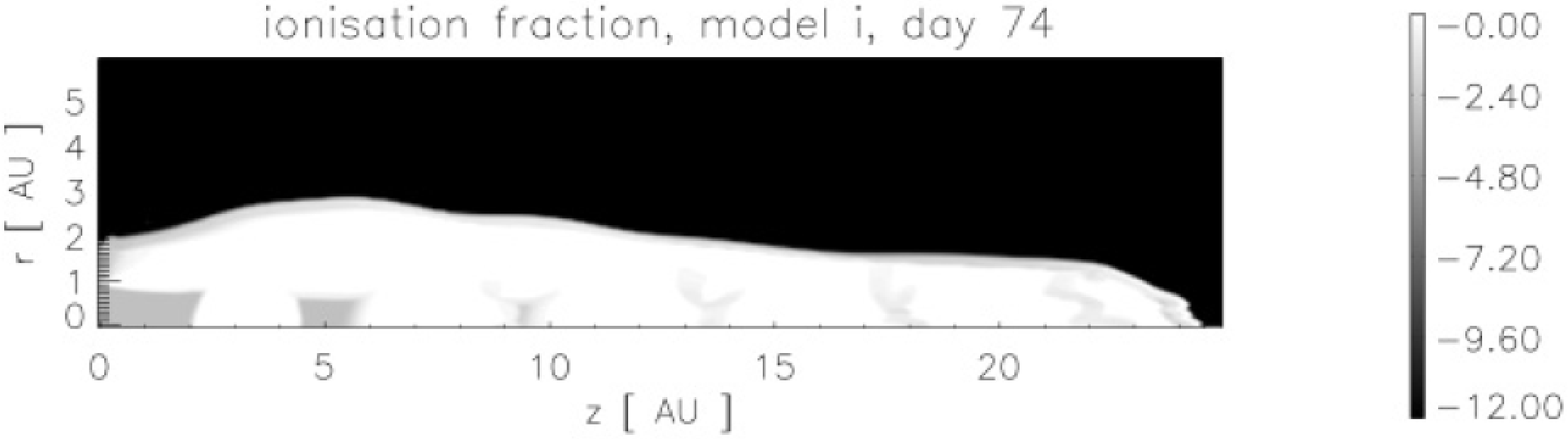}
  \includegraphics[width=8cm]{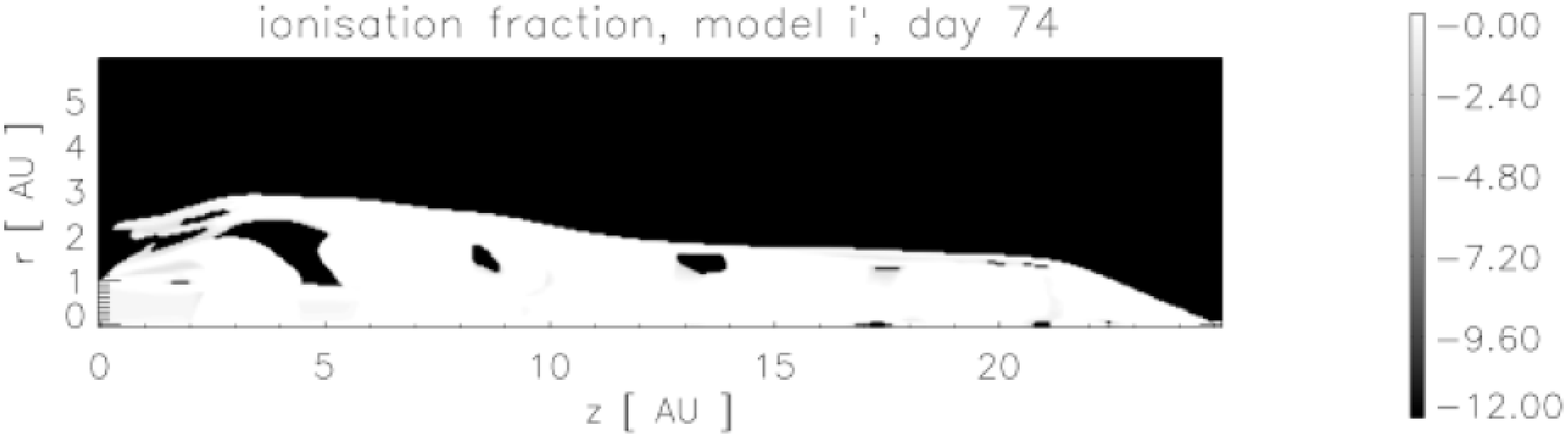}
  \caption{Comparison of temperature (top), density (middle), and 
	ionization fraction (bottom) plots of the old model i (left)
	and model i' (right) at day 74. Besides slight differences in the 
	internal morphology, the dimensions of the jets, positions of the 
	internal shocks, and temperatures are identical in both simulations; 
	differences in the ionization fractions arise in the low-temperature 
	regions of the jet due to the larger equilibration timescales in this 
	temperature range}
  \label{den_comp}
\end{figure*}
\begin{figure*}
  \centering
  \includegraphics[width=8cm]{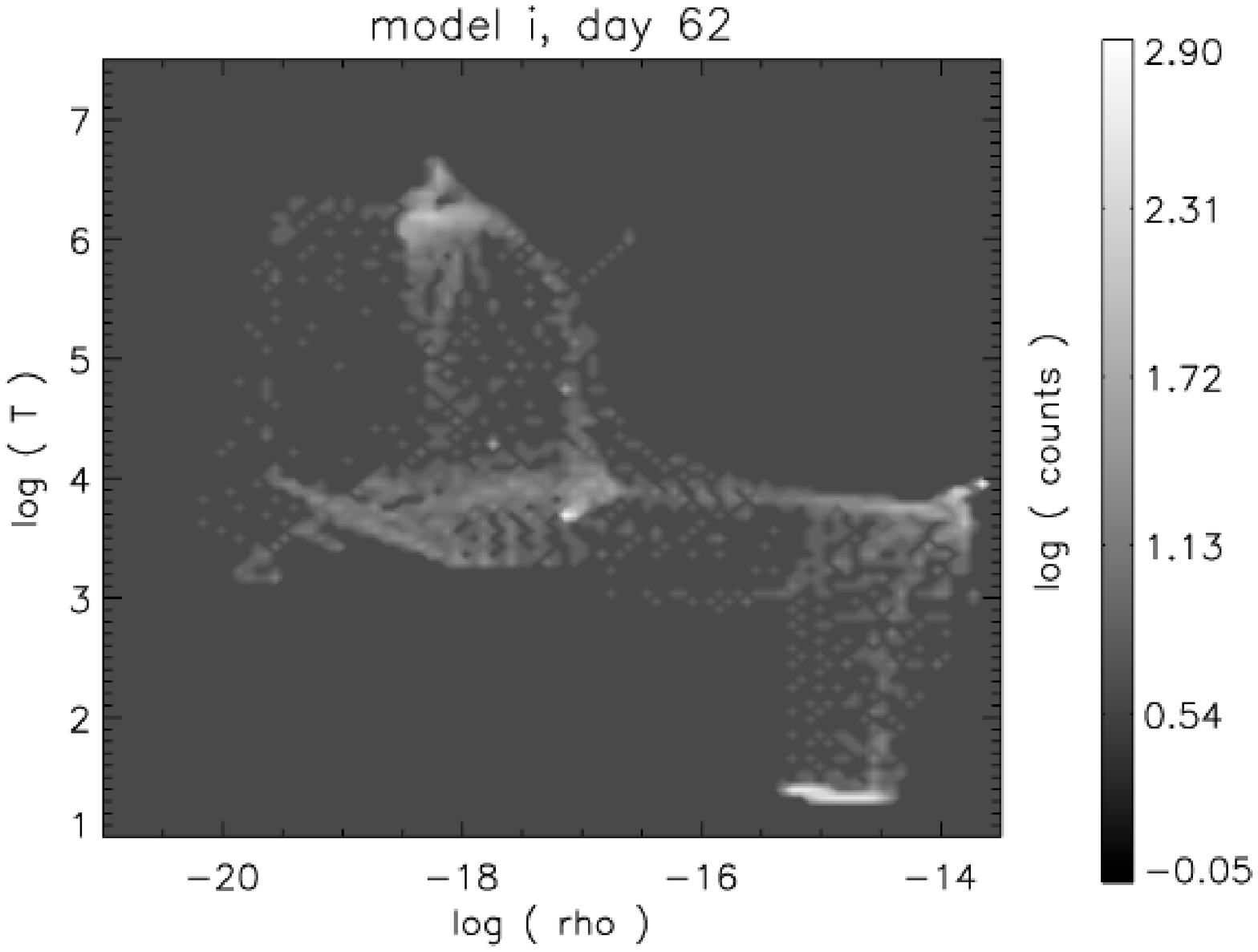}
  \includegraphics[width=8cm]{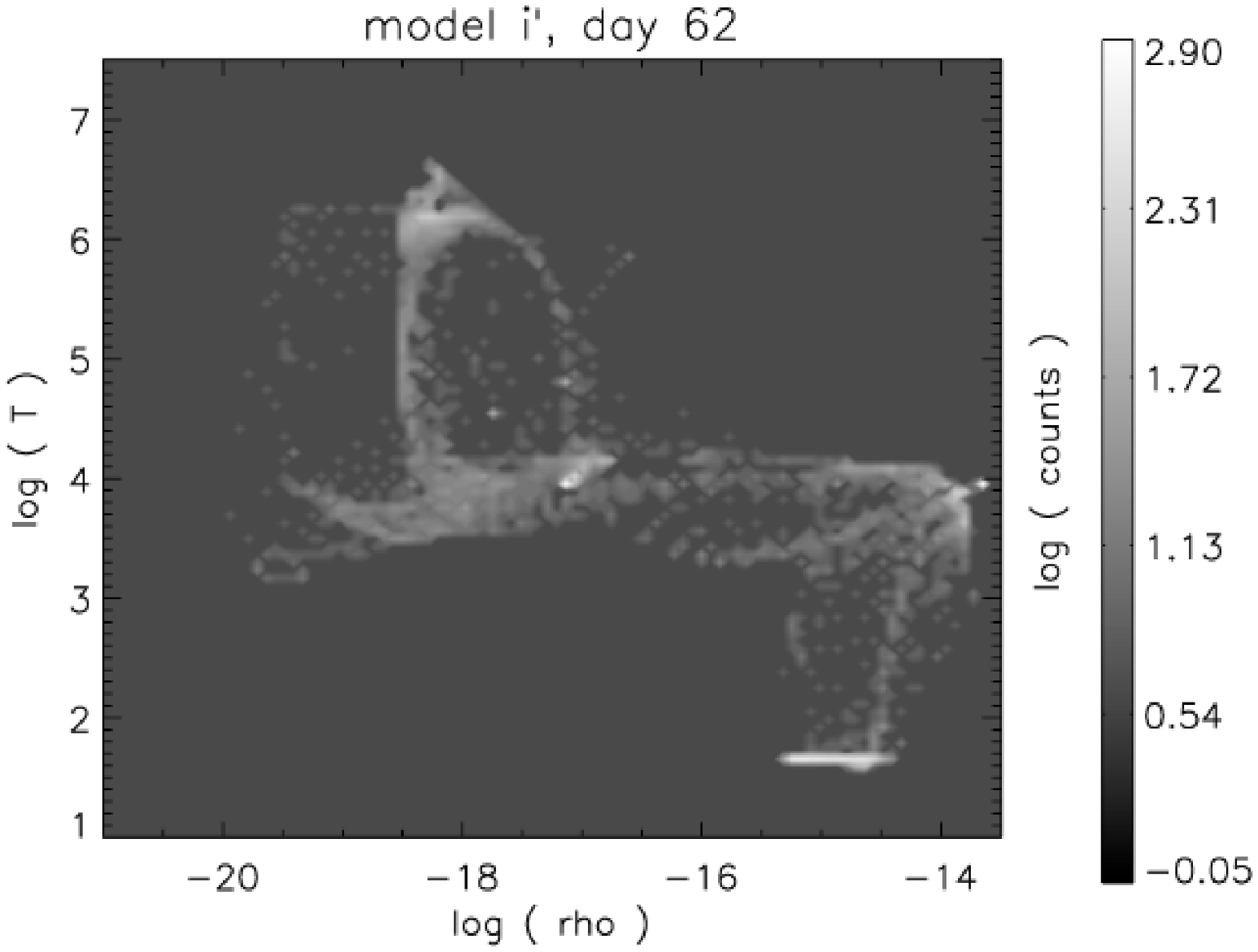}

  \includegraphics[width=8cm]{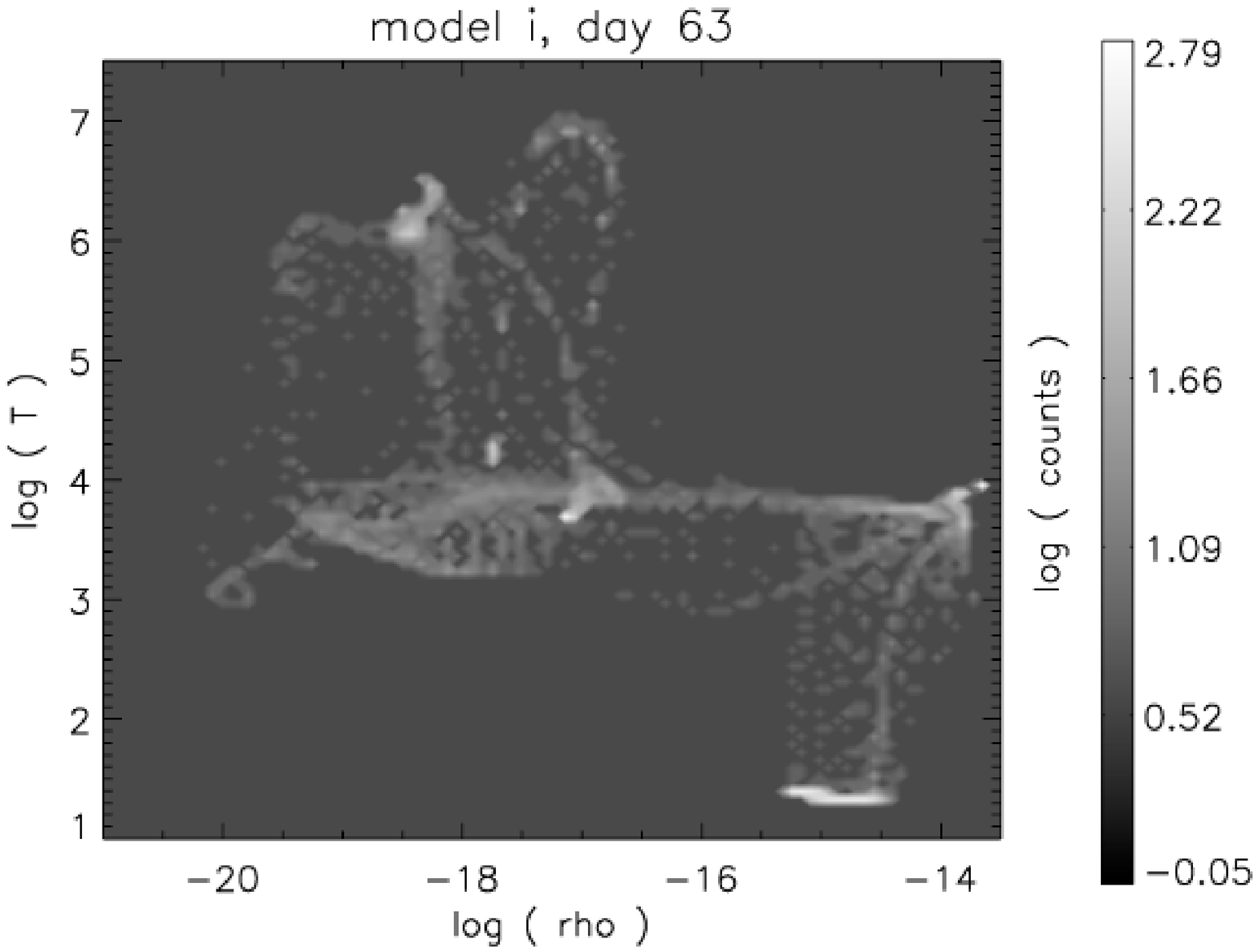}
  \includegraphics[width=8cm]{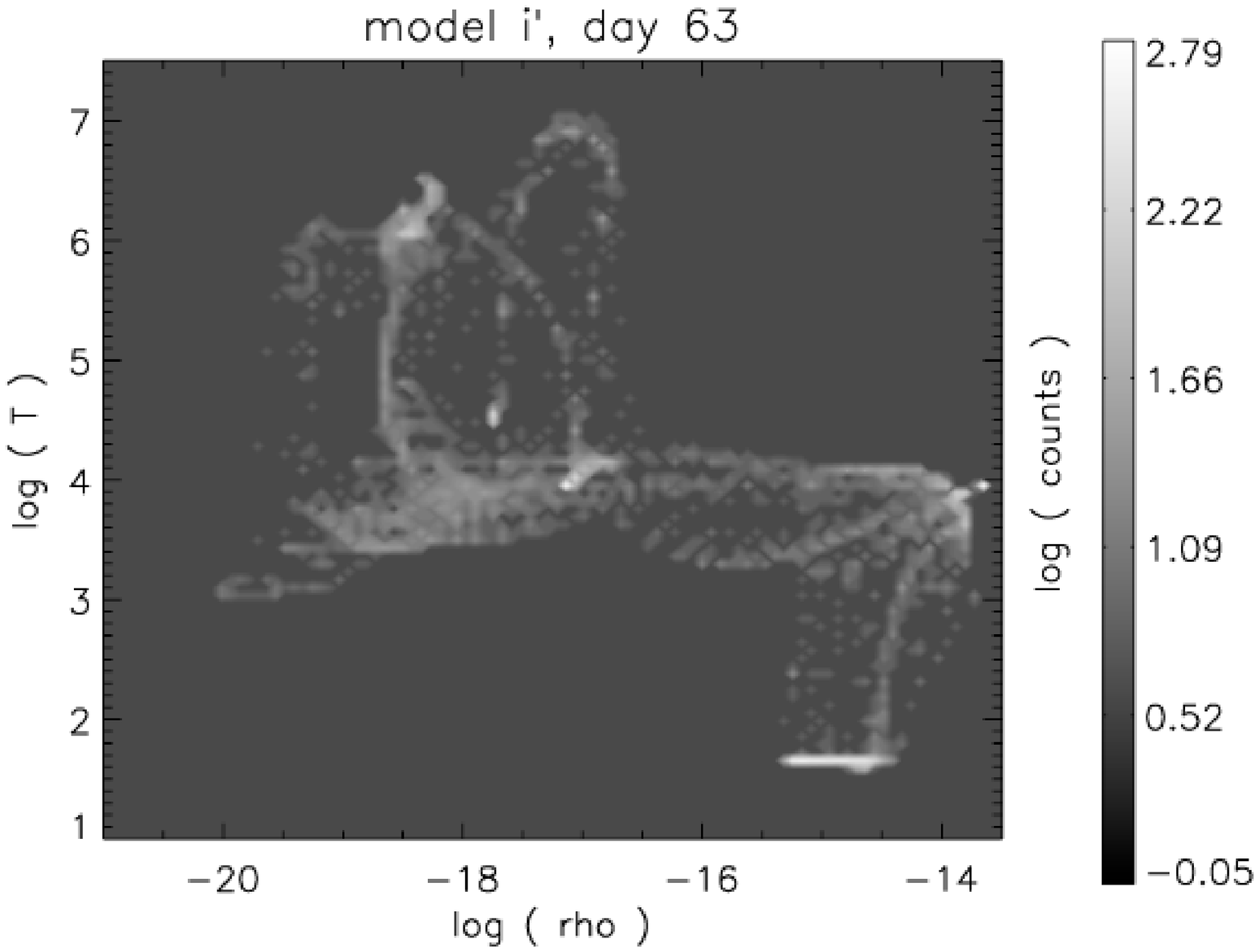}

  \includegraphics[width=8cm]{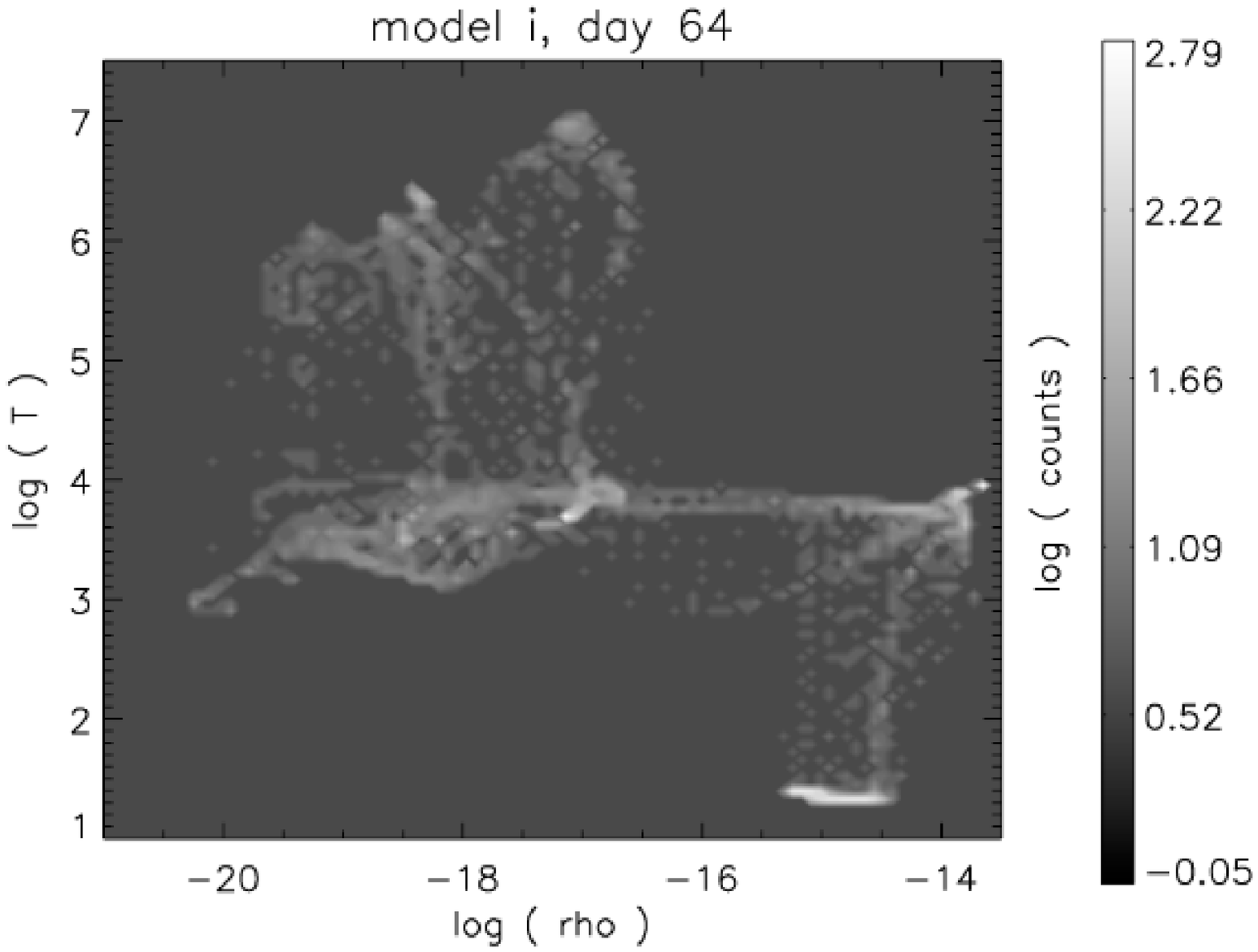}
  \includegraphics[width=8cm]{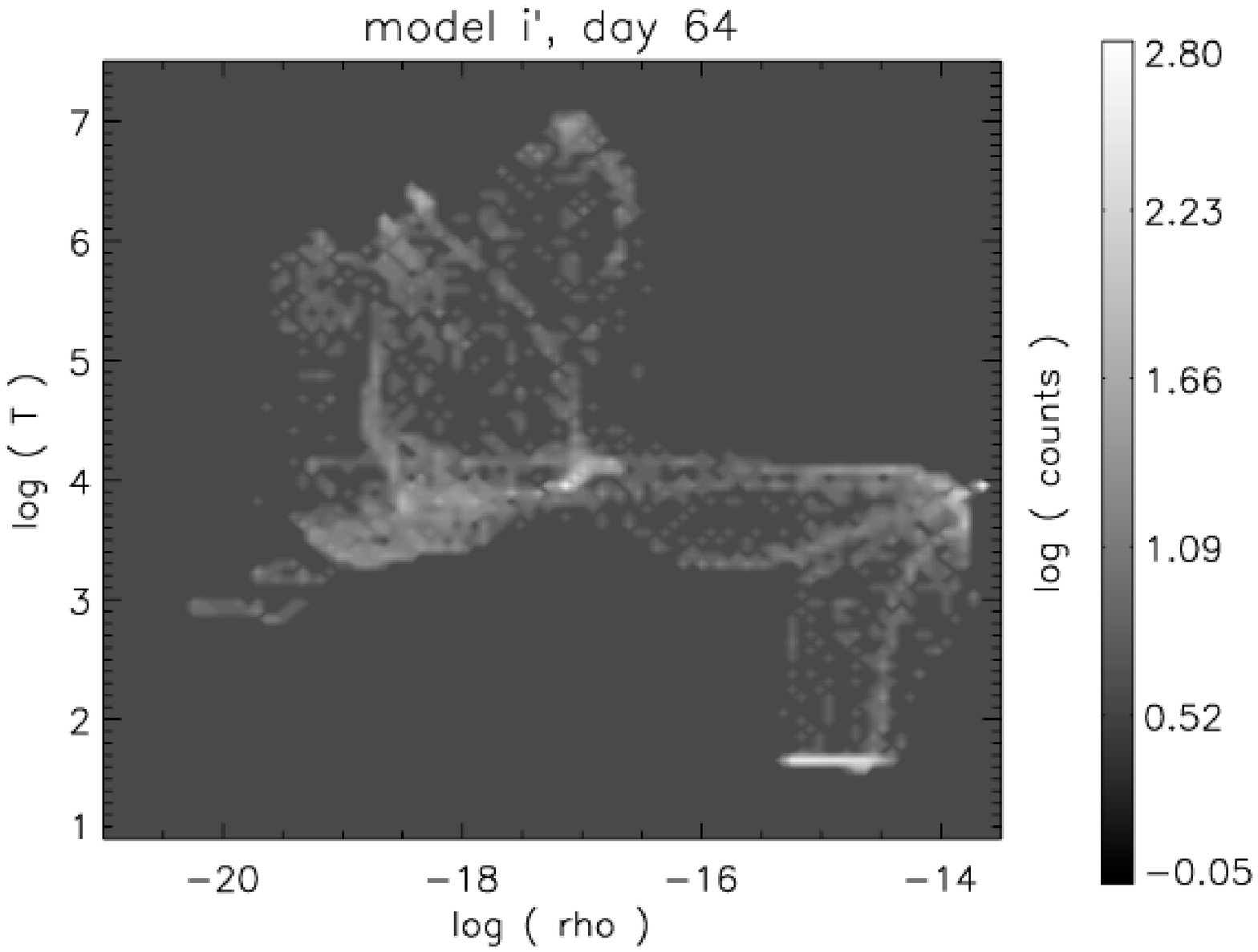}

  \includegraphics[width=8cm]{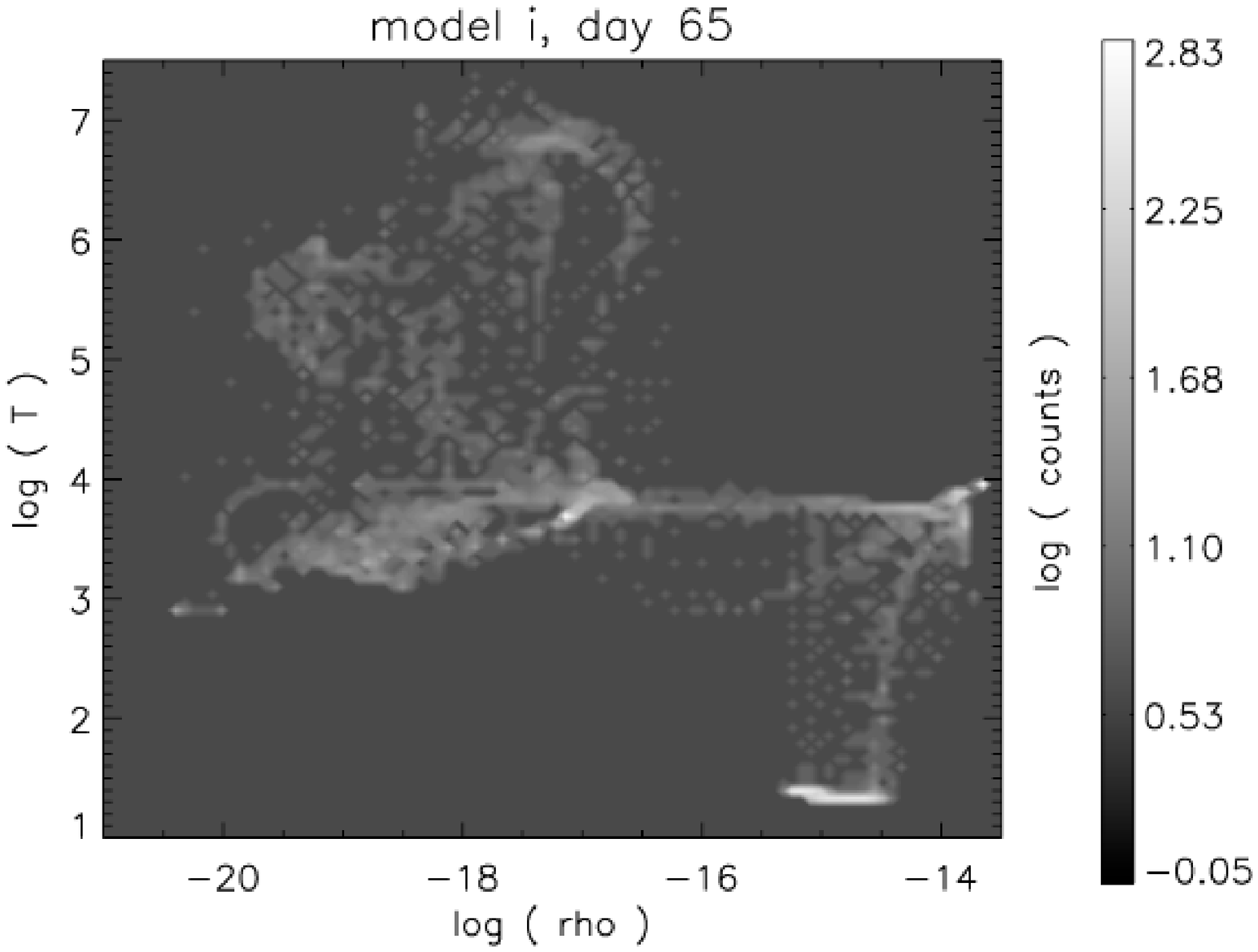}
  \includegraphics[width=8cm]{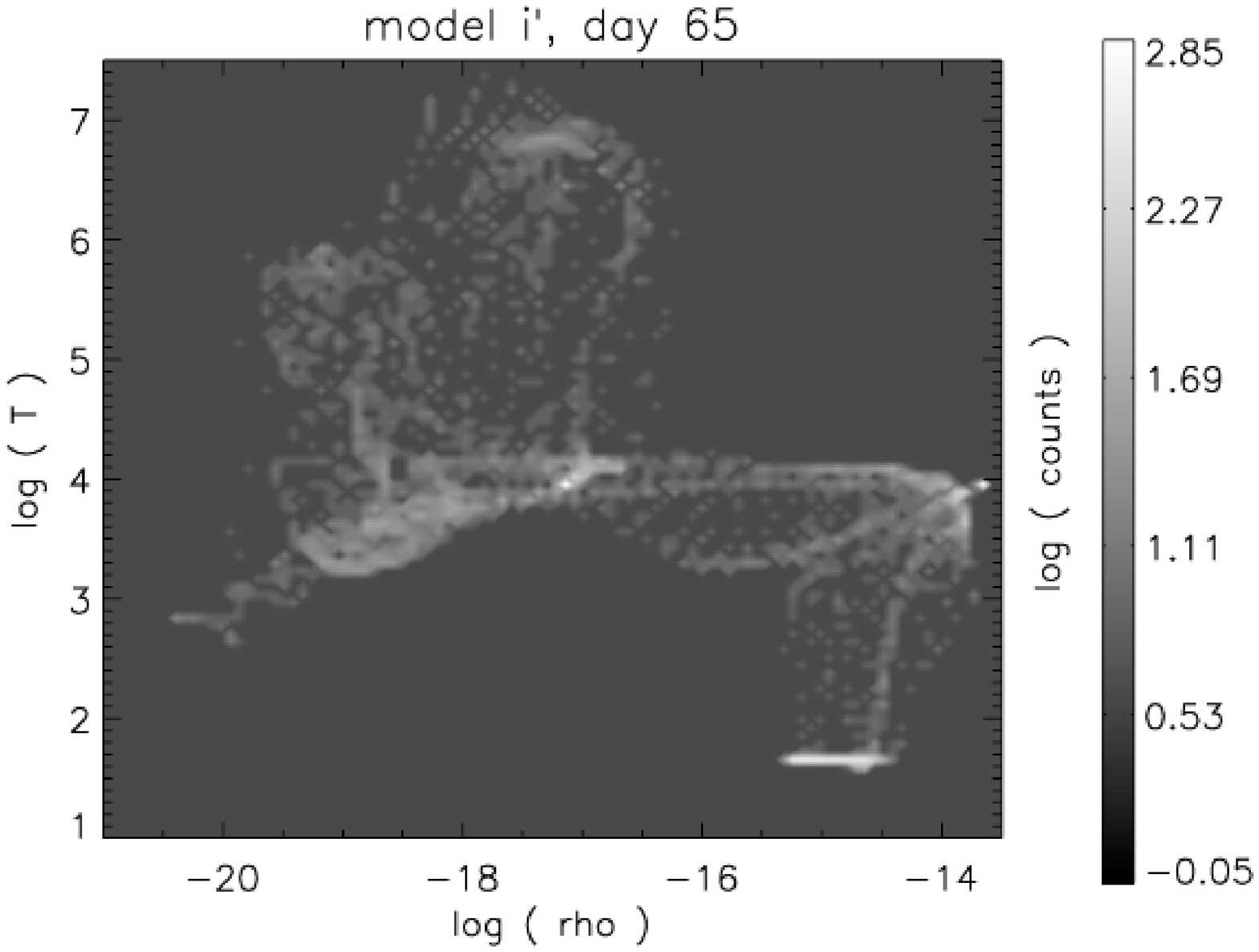}
  \caption{$\rho - T$-histogram of eight consecutive days (day 62 - 69) 
    covering one pulse cycle of model i (left) and model i' (right). 
    The bright bar at very low temperatures and at densities 
    between $10^{-16}$ and $10^{-14}$ g cm$^{-3}$ consists of the external 
    medium. The internal dense knots build up the line at $T = 10^4$ K. The
    new pulse becomes visible as an arc and spreads out along a line of 
    constant temperature}
  \label{rho-t}
\end{figure*}

\begin{figure*}
  \centering
  \includegraphics[width=8cm]{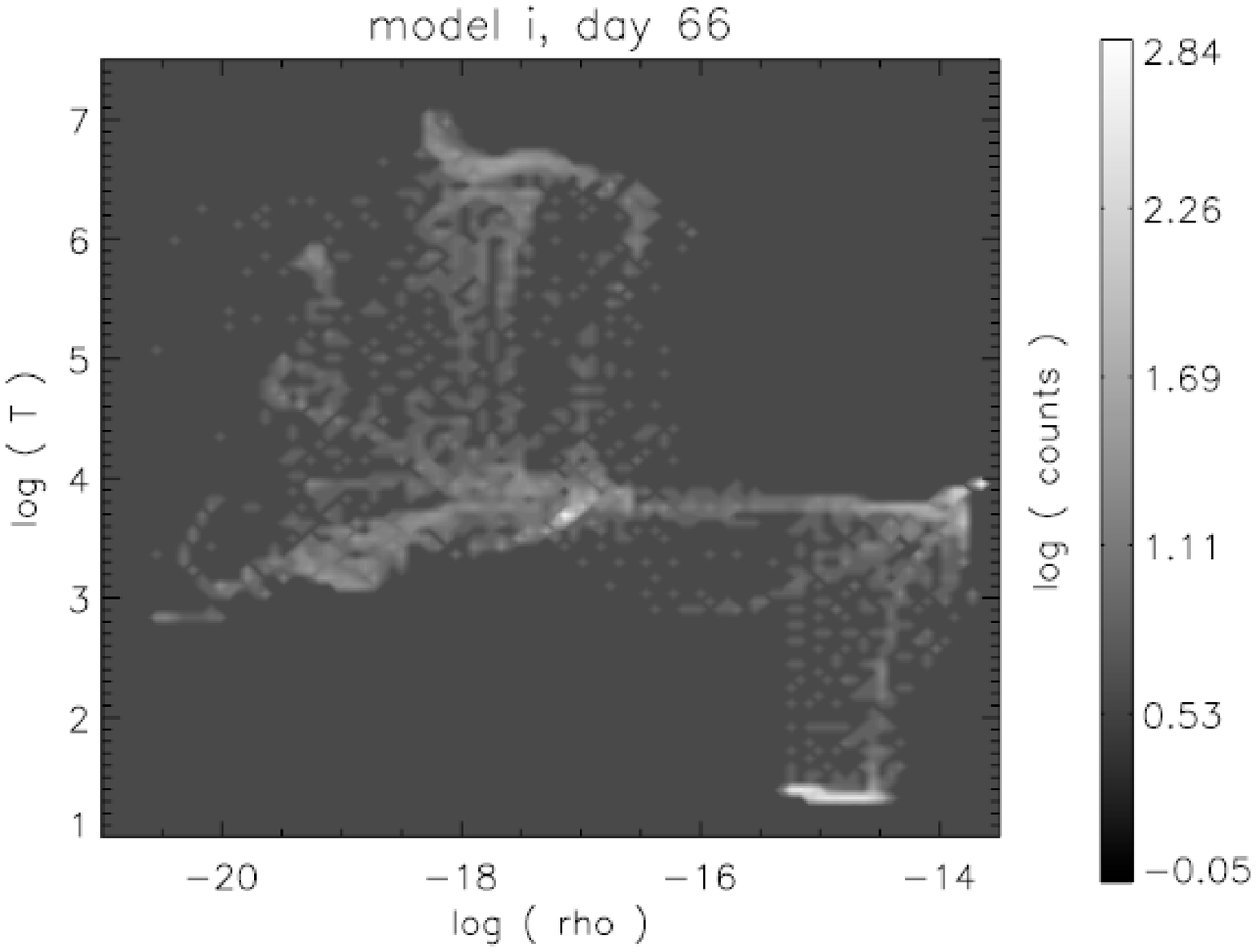}
  \includegraphics[width=8cm]{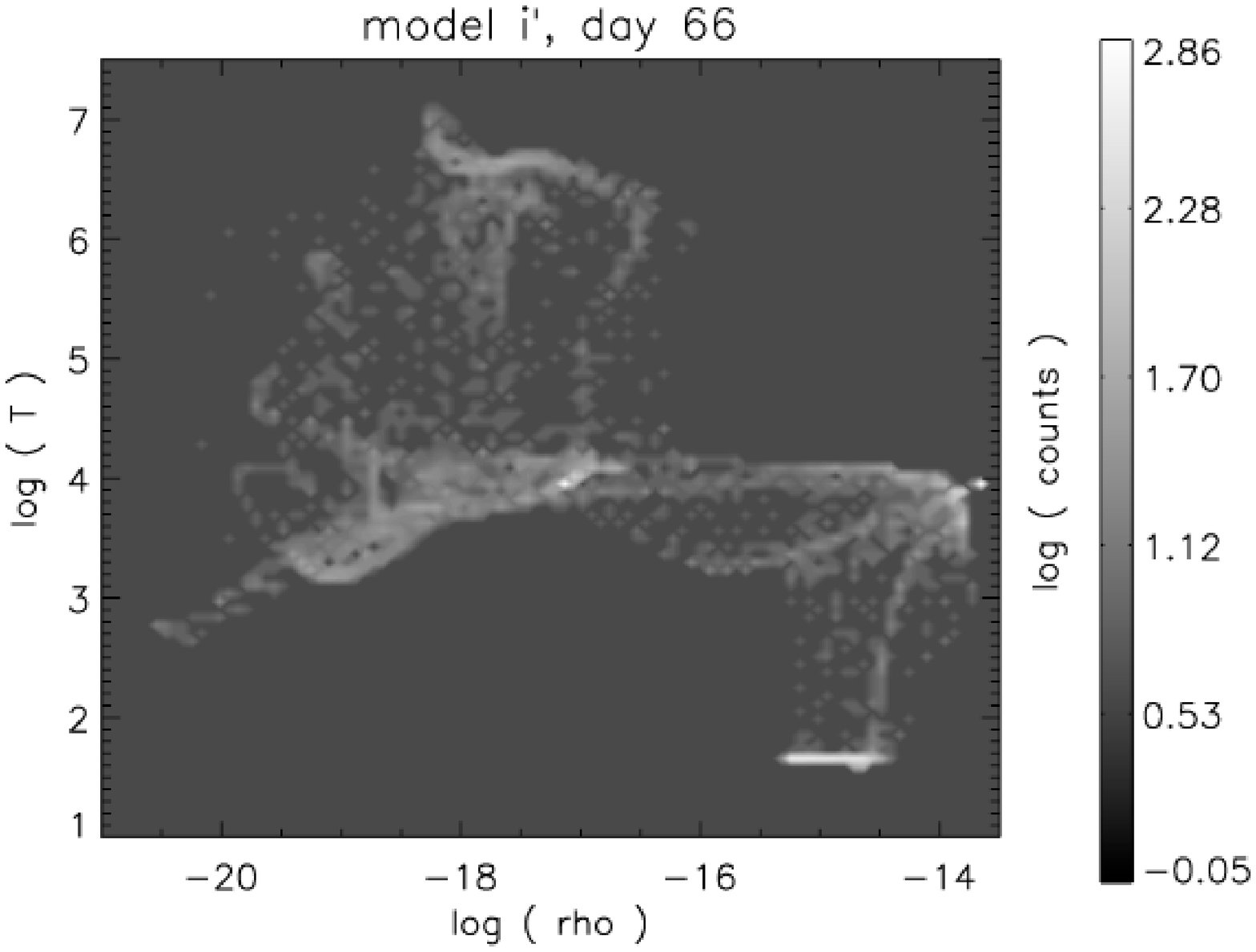}

  \includegraphics[width=8cm]{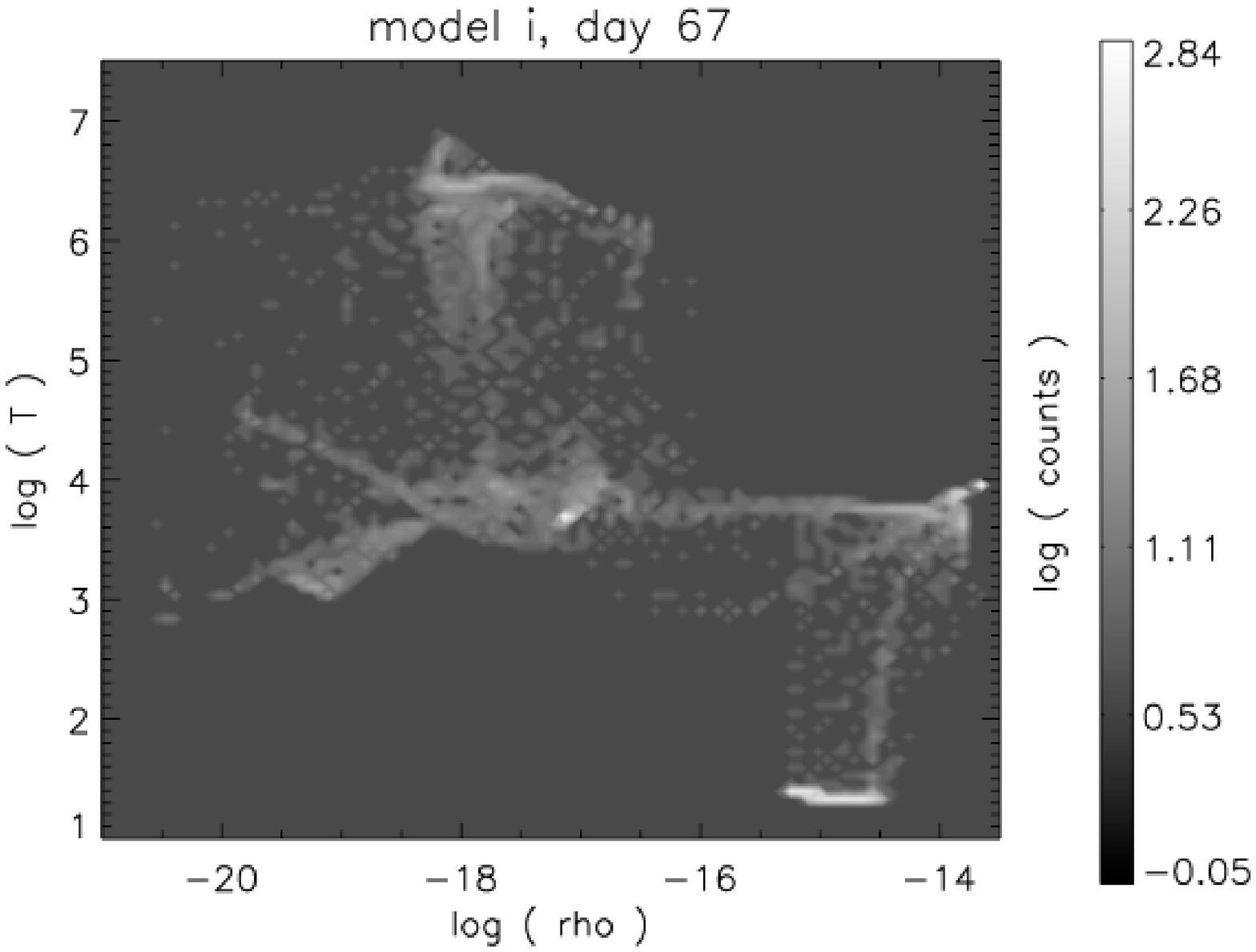}
  \includegraphics[width=8cm]{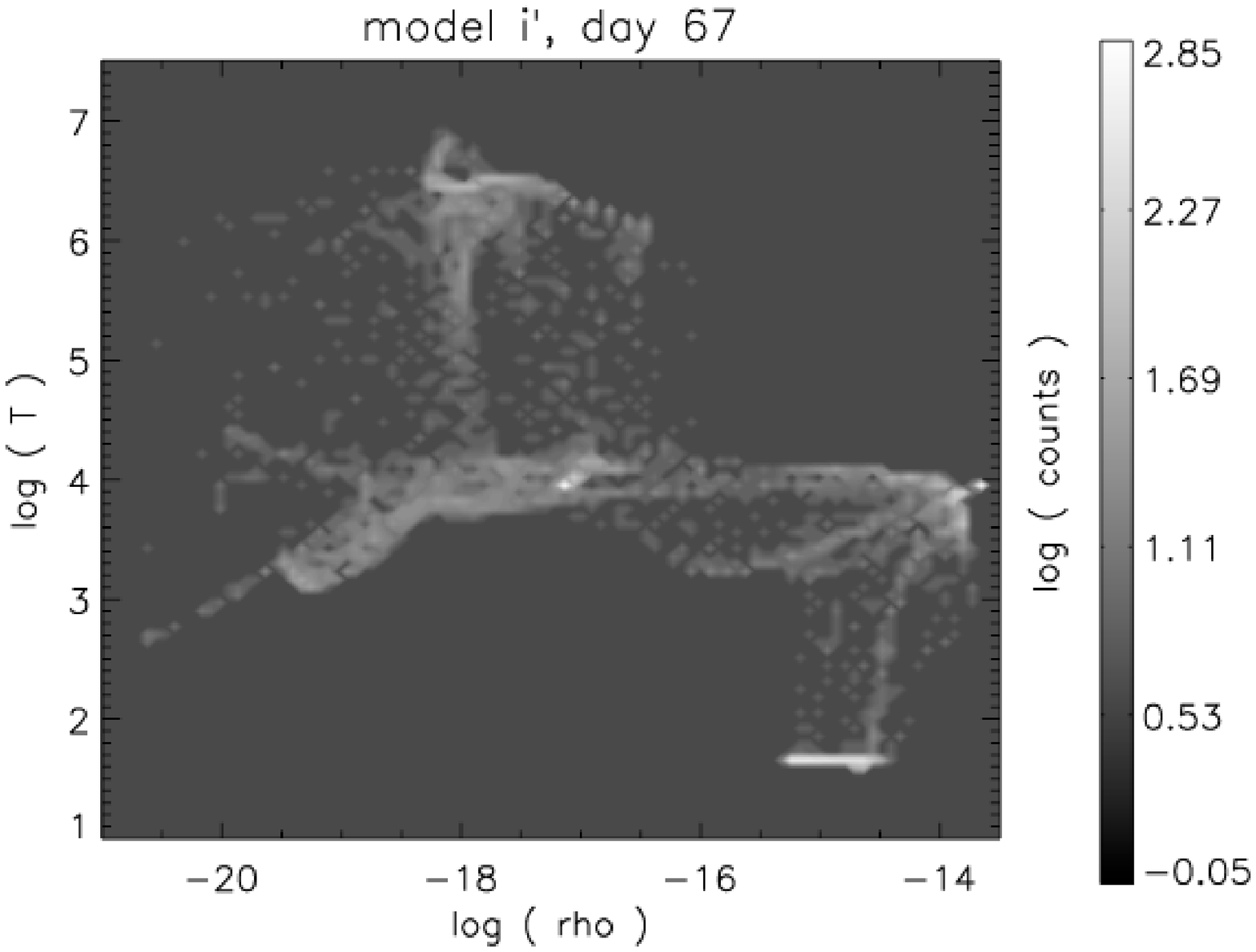}

  \includegraphics[width=8cm]{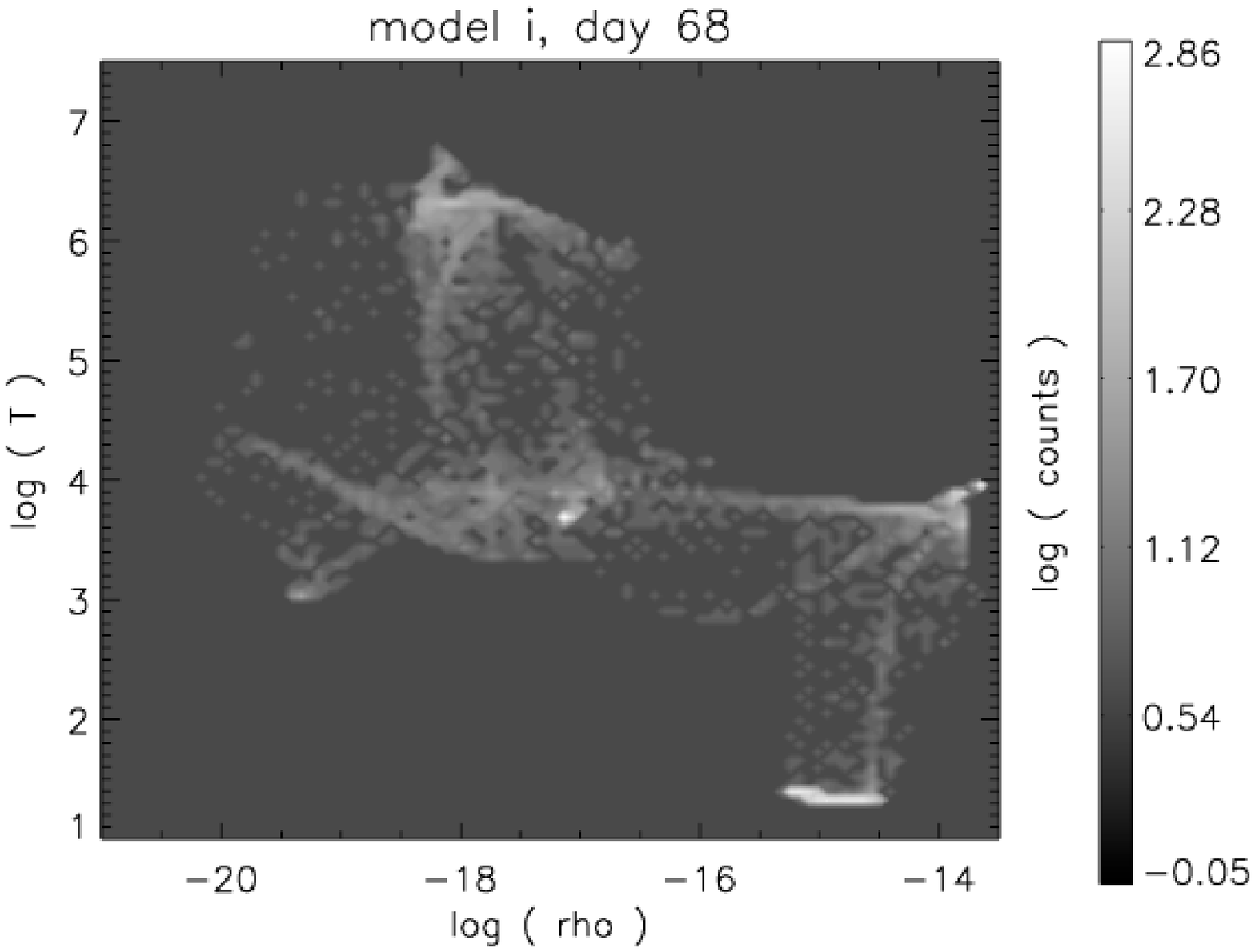}
  \includegraphics[width=8cm]{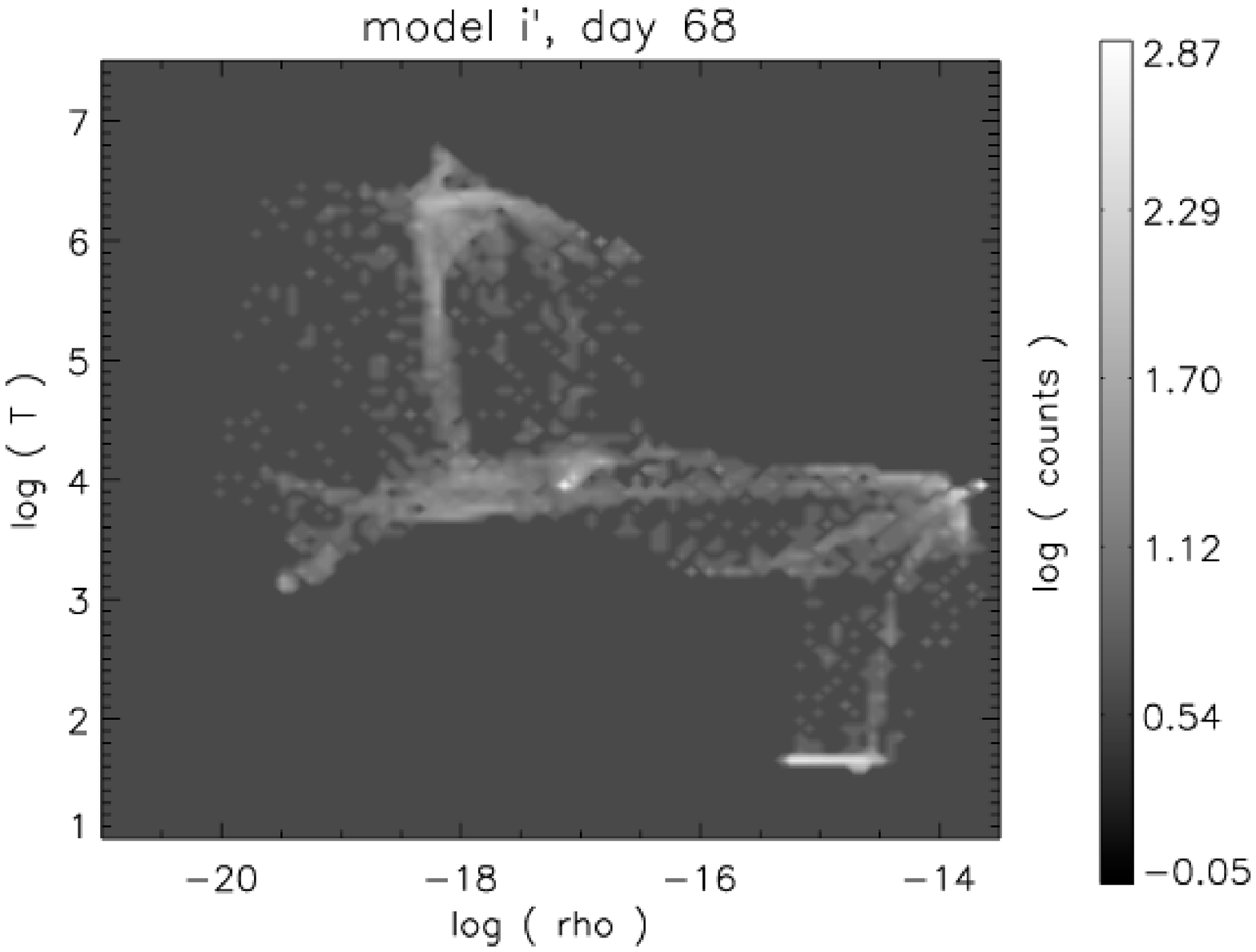}

  \includegraphics[width=8cm]{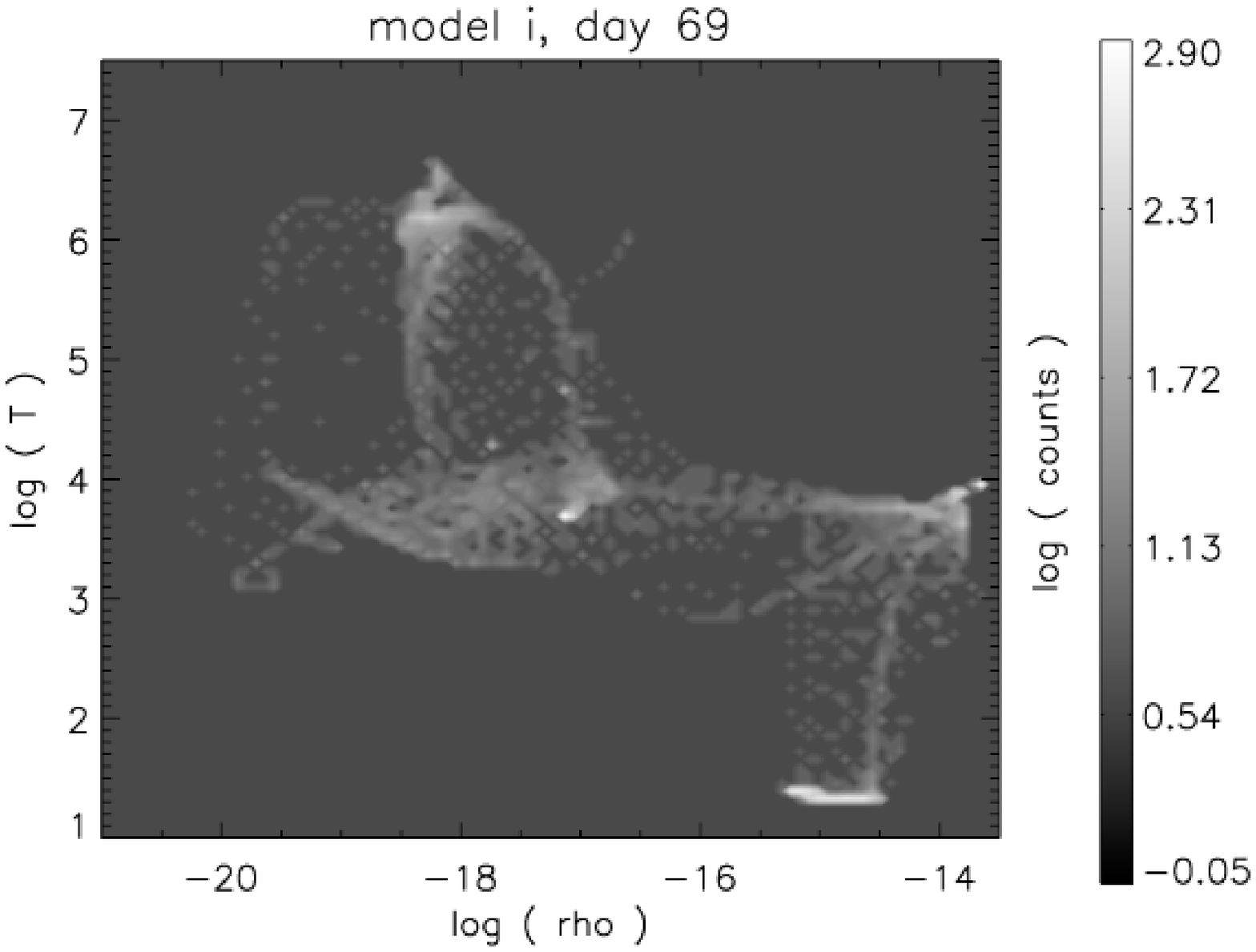}
  \includegraphics[width=8cm]{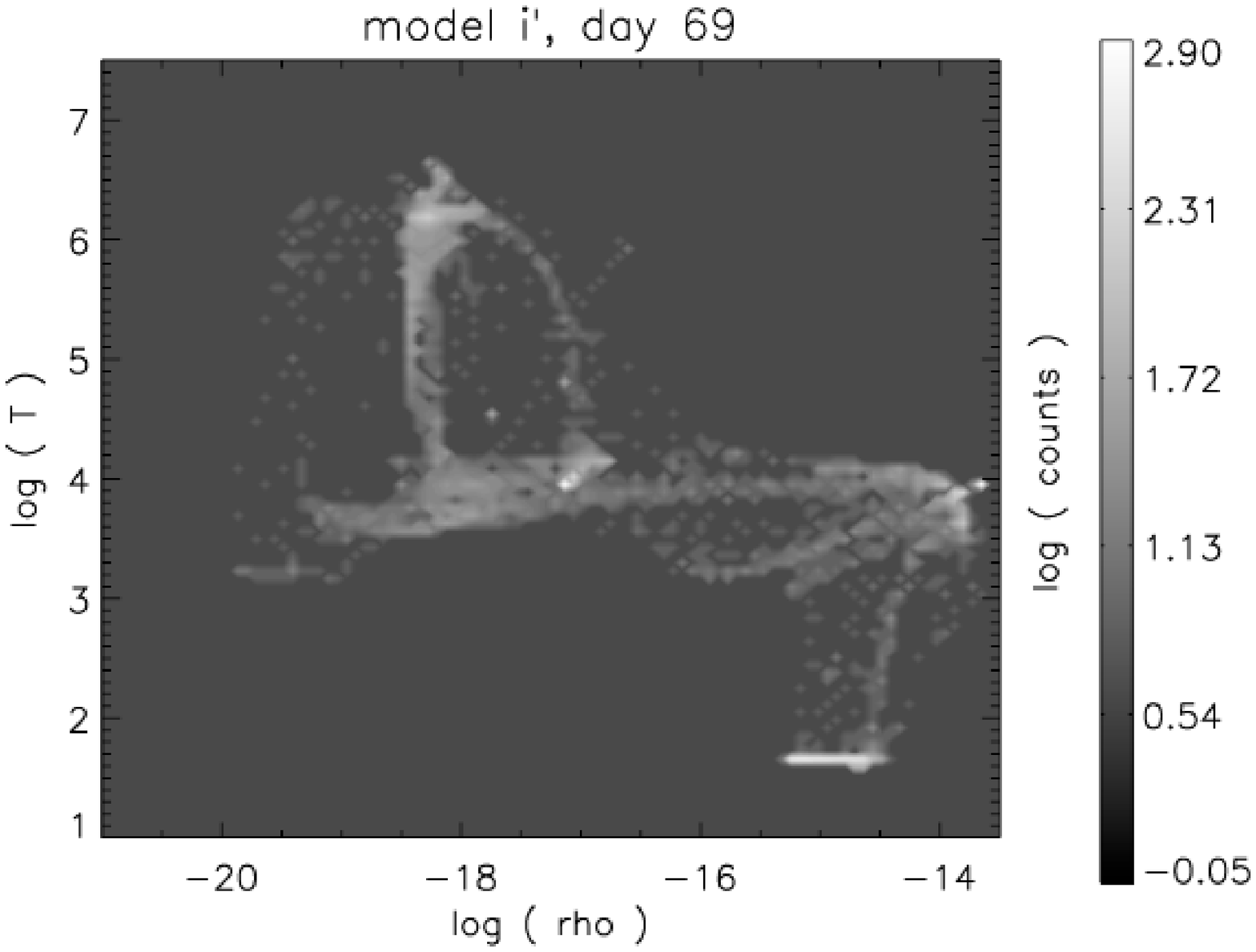}
  \caption{as Fig. \ref{rho-t}, continued. The radiative cooling lowers the 
    temperature of the region of higher densities. Superposed on it, the whole 
    feature moves towards lower densities and lower temperatures, consistent 
    with adiabatic expansion. The slope of the path is approximately 
    $\log ( T ) \sim 2 / 3 \, \log ( \rho )$}
  \label{rho-t2}
\end{figure*}

The general capabilities of the code NIRVANA\_CP have already been described 
in detail in Paper I. The code was written by \citet{ZiY} and modified by 
\citet{Thi} to calculate radiative losses due to non-equilibrium cooling by 
line emission. It solves the equations of ideal hydrodynamics with an 
additional cooling term in the energy equation $\Lambda ( T ; \rho_{i})$ and a 
network of rate equations:
\begin{eqnarray} \label{hydro}
\frac{\partial\,\rho}{\partial\,t} + \nabla\,(\rho\,{\bf v}) &=& 0 \nonumber \\
\frac{\partial\,(\rho\,{\bf v})}{\partial\,t} + \nabla\,
(\rho\,{\bf v}\otimes{\bf v}) &=& - \nabla\,p - \rho\,\nabla\,\Phi \nonumber \\
\frac{\partial\,e}{\partial\,t} + \nabla\,(e\,{\bf v}) &=& - p\,\nabla\,{\bf v}
 + \Lambda ( T ; \rho_{i}) \nonumber \\
p &=& (\gamma - 1)\,e \nonumber  \\
\frac{\partial\,\rho_{i}}{\partial\,t} + \nabla\,(\rho_{i}\,{\bf v}) &=&
\sum_{i=1}^{N_{s}}\,\sum_{j=1}^{N_{s}}\,k_{ij} (T)\,\rho_{i}\,\rho_{j} .
\end{eqnarray}
\noindent
Here $\rho$ is the gas density, $e$ the internal energy, ${\bf v}$ the
velocity, $\gamma$ the ratio of the specific heats at constant pressure and
volume, and $\rho_{i}$ the species densities satisfying
$\rho = \sum_{i=1}^{N_{s}}\,\rho_{i}$ for the total density. $k_{ij}$ are the
rate coefficients for two-body reactions which are functions of the fluid
temperature $T$. They describe electron collision ionization and radiative and
dielectronic recombination processes. When the cooling is solved dynamically
with the full set of non-equilibrium equations, the various ionization states
and concentration densities $\rho_{i}$ of each element are calculated from the
atomic rate equations. They are used explicitly in the cooling functions as
\begin{equation}
\Lambda ( T ; \rho_{i}) = \sum_{i=1}^{N_{s}}\,\sum_{j=1}^{N_{s}}\,e_{ij} (T)\,
\rho_{i}\,\rho_{j} + \Lambda_{BS}
(T)
\end{equation}
\noindent
with $e_{ij}$ the cooling rates from two-body reactions between species $i$ and
$j$, and $\Lambda_{BS}$ the cooling function due to Bremsstrahlung.

Due to constraints set by the available computer resources, our former 
simulation with cooling, presented in Paper I, included a simpler treatment of 
radiative effects. Instead of the full explicit cooling function
$\Lambda ( T ; \rho_{i})$ for all species, we considered only cooling by
hydrogen, together with a general non-equilibrium (NEQ) cooling function
$\Lambda ( T )$ adapted from \citet{SuD93} to account for the cooling by the
heavier elements. This cooling function was derived by calculating
several atomic processes as electron collision ionization, radiative and 
dielectronic recombination, line radiation, and Compton heating
in the temperature range between $10^4$ and $10^{8.5}$ K. For hydrogen 
we solved the atomic network of H\,{\sc i}, H\,{\sc ii}, and e$^{-}$ and 
calculated the cooling due to collisional ionization of H\,{\sc i} and due to 
collisional excitation of hydrogen line emission.

To perform the simulations beyond density balance with cooling, we were forced 
to simplify the treatment of cooling again. We neglected the atomic
network with its rate equations and used the ionization fractions 
of hydrogen as a function of temperature given by \citet{SuD93}, calculated 
by using the recombination, collisional ionization, and charge transfer rates. 
We included Bremsstrahlung and the general cooling function calculated with 
these partial densities. In NEQ models the ionization fraction could be higher 
than in the case of collisional ionization equilibrium (CIE), due to the slow 
process of recombination of hydrogenic and helium-like ions. Only at 
temperatures of about $10^5$ K, however, is the CIE cooling timescale smaller 
than the equilibration timescale for those ions \citep{SuD93}, leading to 
the difference between the CIE and the NEQ cooling curve used. As hydrogen is 
(almost) completely ionized at temperatures higher than 20000 K, the 
differences in the ionization fractions are almost negligible in the 
temperature range of the cooling curve. Small differences should only occur in 
low temperature regions of the jet.

\subsection{The parameters}

The geometrical model is taken over without changes, except for the dimensions
of the computational domain. We choose a two-dimensional slice of the full 
domain and assume axisymmetry. In these new simulations, the dimension in 
axial direction is doubled to 100 AU to be able to study the jet propagation 
after density compensation. The resolution remains unchanged as 20 grid cells 
per AU (jet beam radius).

The hot component is located in the origin of the coordinate frame, therefore
the companion is expanded into a ``Red Giant Ring''. In the 2D integration
domain, only half of the cross-section of this ring on this slice is included.
The binary separation in the models is chosen to 4 AU which is of the order of
the estimated separations of 3.3 - 5.2 AU. The density of the red giant is set
to $2.8 \cdot 10^{-5}$ g cm$^{-3}$ and its radius to 1 AU.

Surrounding the red giant, a stellar wind is implemented. The wind has a
constant velocity of $v=10$~km s$^{-1}$, a gas temperature of $T=50$ K, and a 
mass loss of 10$^{-6}$ $\mbox{M}_{\odot}$ yr$^{-1}$. The density of the red 
giant wind at the surface of the star is then $2.2 \cdot 10^{-14}$ g cm$^{-3}$.
The density of the external medium is given by a $1/r^2_{\rm{rg}}$-law for a
spherical wind where $r_{\rm{rg}}$ is the distance from the center of the
red giant. The density of the red giant wind near the jet nozzle is about 200
times higher than the initial jet density. At a distance of 50 AU from the
symbiotic system, i.e. at the end of the integration domain of the former 
simulations, the wind density is about equal to the jet density at the nozzle.

To account for the counter-jet and the other part of the jet, respectively, 
the boundary conditions in the equatorial plane and on the jet axis are set to 
reflection symmetry. On the other boundaries, outflow conditions are chosen.

The jet is produced within a thin jet nozzle with a radius of 1 AU. The 
initial velocity of the jet is chosen to 1000 km s$^{-1}$ or 0.578 AU d$^{-1}$ 
and its density is set to $8.4 \cdot 10^{-18}$ g cm$^{-3}$ (equal to a 
hydrogen number density of $5 \cdot 10^6$ cm$^{-3}$). These parameters lead to 
a density contrast $\eta$ of $5 \cdot 10^{-3}$, a Mach number of $\approx 60$ 
in the nozzle and a mass loss rate of $\approx 10^{-8}$ 
$\mbox{M}_{\odot}$ yr$^{-1}$. Repeatedly each seventh day, the velocity and 
density values in the nozzle are changed to simulate the jet pulses which are 
seen in the observations of MWC 560.

The pulse parameters of the first simulation were chosen as in the old model i 
with cooling of Paper I. The nomenclature is now model i' -- mention of the 
cooling will be omitted hereafter. A second simulation with parameters as 
in the adiabatic model iv -- the model with the highest density pulses -- was 
also performed and is now named model iv'.

Table \ref{pulses} lists the main parameters of the models: the
model number, the density of the pulse in the nozzle n$_{\rm{pulse}}$ in
cm$^{-3}$, the velocity of the pulse in the nozzle v$_{\rm{pulse}}$ in cm
s$^{-1}$, the mass loss during the pulse $\dot M$ in g s$^{-1}$ and in
$\mbox{M}_{\odot}$ yr$^{-1}$, and the kinetic jet luminosity during
the pulse in erg s$^{-1}$. Each pulse lasts for one day. 

\begin{table*}
\caption{Parameters of the jet pulses: the model number, the density of the
pulse in the nozzle n$_{\rm{pulse}}$ in cm$^{-3}$, the velocity of the
pulse in the nozzle v$_{\rm{pulse}}$ in cm s$^{-1}$, the mass outflow during
the pulse $\dot M$ in g s$^{-1}$ and in $\mbox{M}_{\odot}$ yr$^{-1}$, and the
kinetic jet luminosity during the pulse in erg s$^{-1}$.} \label{pulses}
\begin{flushleft}
\begin{tabular}{lcccccc}
\hline
model & n$_{\rm{pulse}}$ [cm$^{-3}$] & v$_{\rm{pulse}}$ [cm s$^{-1}$] &
$\dot M$ [g s$^{-1}$] & $\dot M$ [$\mbox{M}_{\odot}$ yr$^{-1}$] &
L$_{jet}$ [erg s$^{-1}$] \\
\hline
i' & $1.25 \cdot 10^6$ & $2.0 \cdot 10^8$ & $2.94 \cdot 10^{17}$ &
$4.66 \cdot 10^{-9}$ & $5.88 \cdot 10^{33}$ \\
iv' & $1.0 \cdot 10^7$ & $2.0 \cdot 10^8$ & $2.35 \cdot 10^{18}$ &
$3.73 \cdot 10^{-8}$ & $4.70 \cdot 10^{34}$ \\
\hline
\end{tabular}
\end{flushleft}
\end{table*}

\section{Validating the cooling treatment} \label{sec_valid}

By the aforementioned simplifications of the cooling treatment, we achieved 
a reduction in the required computational time by a factor of 8--10 with 
respect to the old simulation with cooling. This was a necessary condition to 
be able to enlarge the computational domain. Before investigating the new 
simulations, however, we now have to validate the ability of our approximated 
cooling treatment to describe the real properties inside jets of symbiotic 
stars. 

The first way is a comparison of the plots of density, temperature, and -- 
more important -- the ionization fraction $X = n_{e} / n_{H}$, the variables 
that determine the observable emission. In Fig. \ref{den_comp} we show 
temperature, density, and ionization fraction plots from models i and i' 
at day 74, the last simulated time step in the old run. Apart from the slight 
differences in the internal structure, such as in the shapes of vortices, the 
overall results for the bow-shock sizes of the jets, positions of the internal 
shocks, and density values are identical in both simulations. Therefore the 
previously derived properties, such as the jet structure, the bow-shock 
geometry and evolution, the internal jet structure, and pulse evolution are 
consistent in both simulations. The same also holds for the temperature. 
As mentioned in Sect. \ref{sec_model}, the recombination time, 
thus the equilibration timescale, is larger than the cooling timescale
at smaller temperatures ($10^4$ - $10^5$ K), which leads to an underestimation 
of the ionization fraction in those jet regions -- the holes in Fig. 
\ref{den_comp}. As these parts are also of low density, the effects on the 
cooling and the jet emission should be small. This result seems to legitimate 
the use of a simplified treatment of the cooling inside jets in symbiotic 
stars and makes it possible to enlarge the numerical investigations with 
reduced computational constraints.

In a two-dimensional histogram of the density and the temperature as given in 
Figs. \ref{rho-t} and \ref{rho-t2}, one can compare the results of both runs 
in greater detail. They show the physical conditions of the gas
irrespective of their position. 

The bright bar at very low temperatures and at densities between $10^{-16}$ 
and $10^{-14}$ g cm$^{-3}$ consists of the external medium and the internal 
dense knots build up the line at $T = 10^4$ K. Each new pulse becomes visible 
as an arc in the $\rho-T$-plane with its tip at $\rho=10^{-17}$ g cm$^{-3}$ 
and $T=10^{7.5}$ K. On the third day after the pulse, this tip spreads out 
along a line of constant temperature between densities of $10^{-18\ldots-16}$ 
g cm$^{-3}$. Then the radiative cooling comes into play by lowering the 
temperature of the region of higher densities. Superposed on the radiative 
cooling, the whole feature moves towards lower densities and lower 
temperatures. The slope of the path is approximately 
$\log ( T ) \sim 2 / 3 \, \log ( \rho )$, consistent with an adiabatic 
expansion.

The density and temperature maps were very similar (Fig. \ref{den_comp}), and 
the statistical investigation also shows no qualitative differences. The shape 
and position of the pulse structure, of the jet knots, the jet beam, and the 
external medium in the $\rho-T$-plane are identical in both runs. Both cooling 
approaches are therefore similar and seem to be equally reasonable physically; 
the practical differences, however, are huge as mentioned before.

\section{Jet structure} \label{sec_struct}

\subsection{Bow-shock geometry and evolution}

\begin{figure}
  \resizebox{\hsize}{!}{\includegraphics{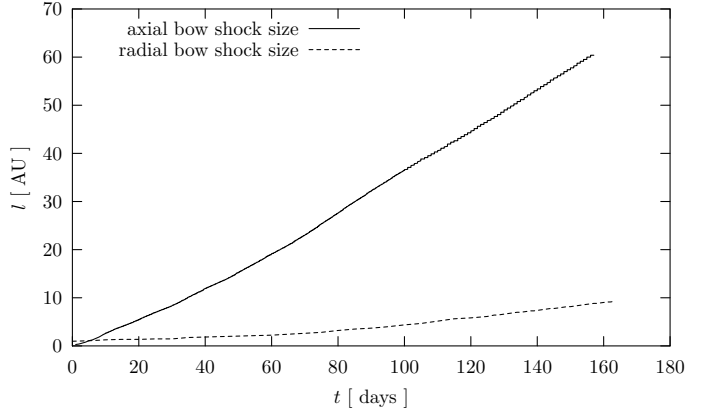}}
  \caption{Size of the bow-shock in axial and radial direction for model i'.
    After a phase of constant radial extent and therefore acceleration of the 
    axial bow-shock, the jet head grows, leading to a constant propagation 
    velocity of the jet}
  \label{NV3.1bow}
\end{figure}

One result of the small-scale simulation with cooling presented in Paper I was 
that the radial extent of the lateral jet bow-shock in model i remained 
constant, which was a significant difference to the adiabatic models where both
the radial and the axial extent increased with a constant aspect ratio up to 
an axial extent of about 35--50 AU, depending on the model parameters.

At the beginning, the radial extent of the jet in model i' and the cross 
section of the jet stay constant as in model i. This again leads to a faster 
propagation velocity than in the adiabatic jet, which even accelerates (Fig. 
\ref{NV3.1bow}). The radial extent and, therefore, the jet head area are not 
able to compensate for the decreasing local density contrast as in the 
simulations without cooling. After day 70, however, the axial velocity 
suddenly seems to remain constant. The radial extent and the jet head area 
grow and the acceleration stops. This is a new behavior not seen in model i as 
only 74 days were simulated in that run. The then constant 
velocity of the jet head is about 740 km s$^{-1}$.

\begin{figure}
  \resizebox{\hsize}{!}{\includegraphics{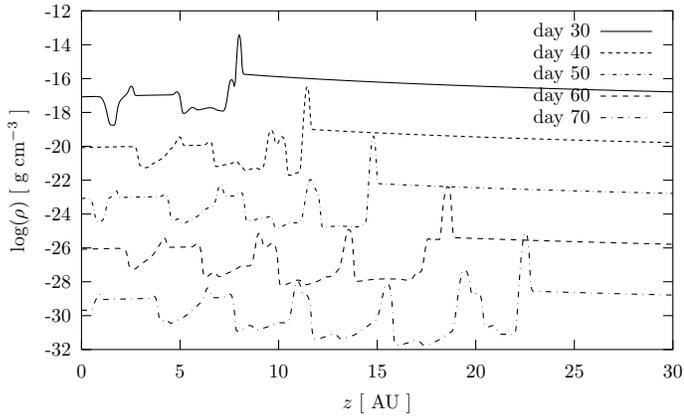}}
  \caption{Slices of density along the jet axis for model i' at days 30, 40, 
    50, 60, and 70, respectively. The last four lines are shifted downwards by 
    3, 6, 9, and 12 orders of magnitude, respectively, for clarity. After day 
    70, the density of the first internal shock behind the bow-shock is larger 
    than the density of the surrounding medium, which then leads to the 
    substantial change in kinematics}
  \label{trans}
\end{figure}

The aspect ratio $r / z$ decreases noticeably in the first 20 days due to 
the constant radial extent $r$. After 70 days the aspect ratio itself remains 
constant, as seen in the {\em initial} phase of the adiabatic models.
First, it is surprising that this transition, which is naively expected due to 
the density balance of the jet and its surrounding medium at a distance of 50 
AU, already occurs on day 70 when the jet has propagated only 20 AU. When 
investigating slices of density along the jet axis, however, one realizes that 
this transition is indeed caused by a density balance (Fig. \ref{trans}). 
\begin{figure*}
  \includegraphics[width=8cm]{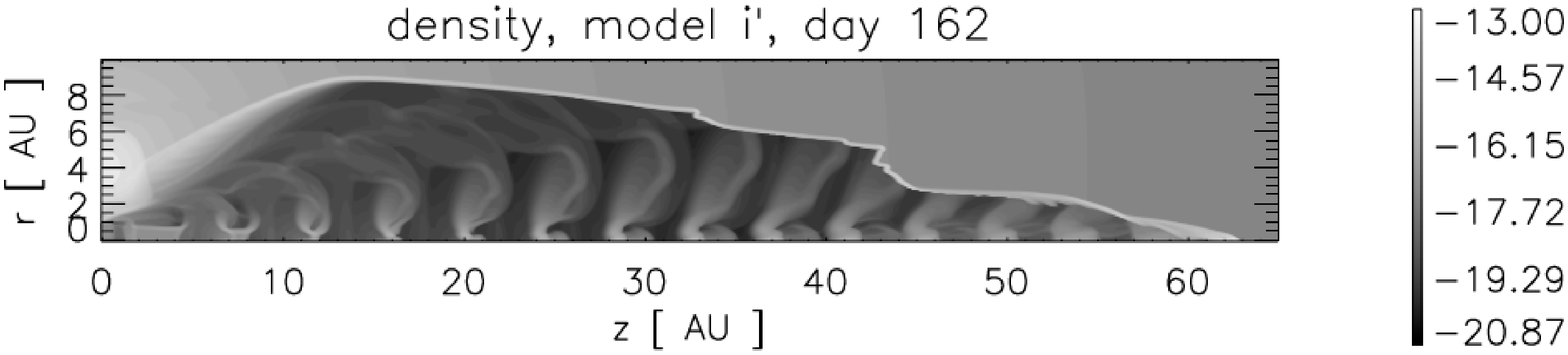}
  \includegraphics[width=8cm]{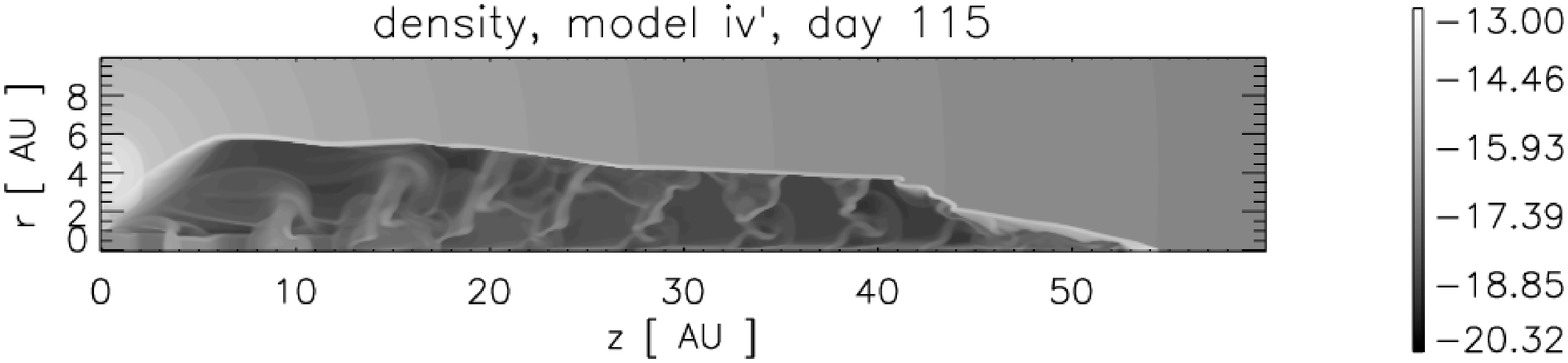}

  \includegraphics[width=8cm]{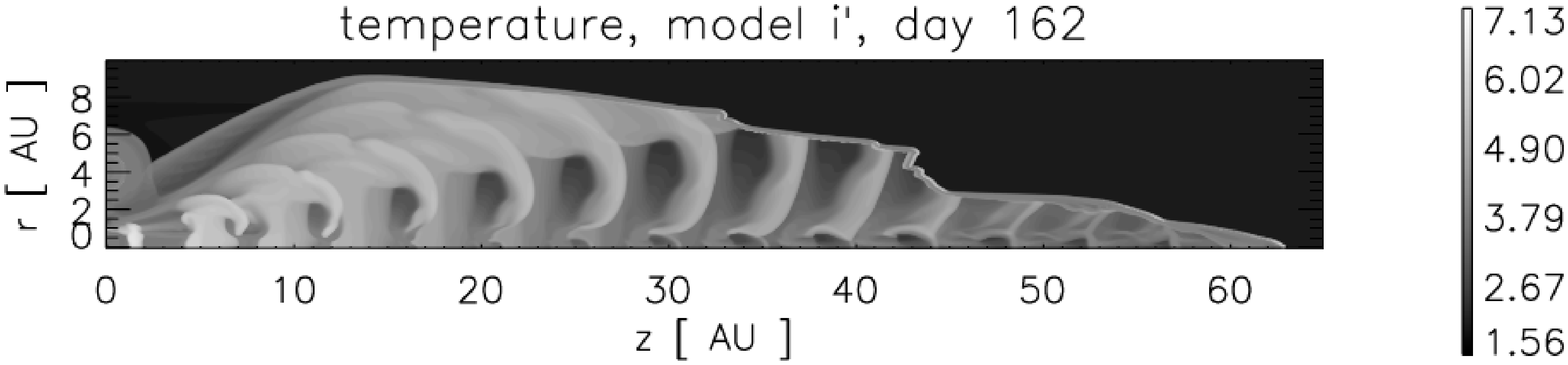}
  \includegraphics[width=8cm]{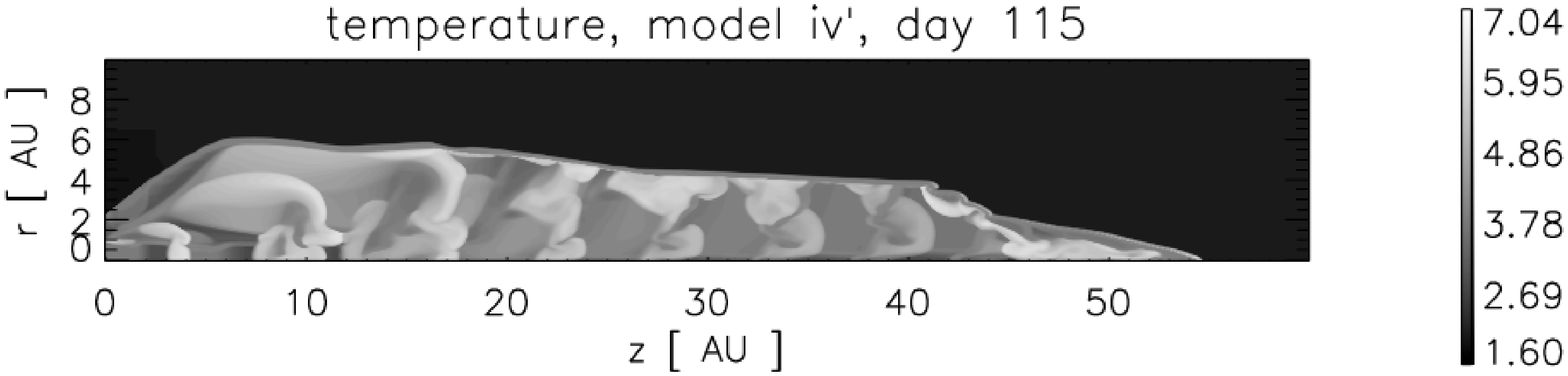}

  \includegraphics[width=8cm]{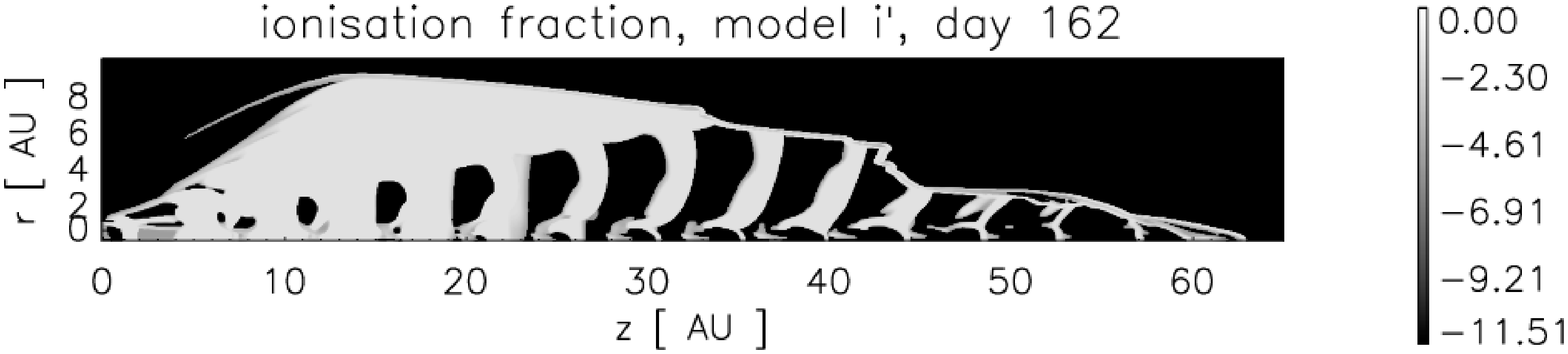}
  \includegraphics[width=8cm]{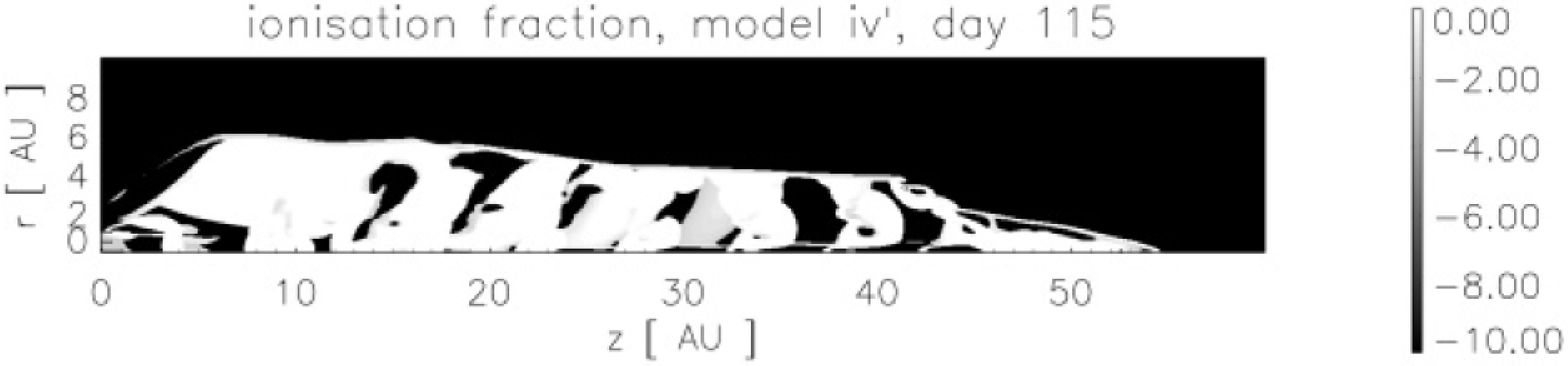}
  \caption{Logarithm of density for model i' on day 162 and for model iv' on 
 	day 115. The extent of the knots in the jet beam is enlarged with 
	respect to the former model i with cooling. No cocoon and no backflow 
	are present; in model iv' the internal structure is more complicated 
	than in model i' and the knots are drifting away from the jet axis}
  \label{NV3.1den}
\end{figure*}

After day 70, the density of the first internal shock behind the bow-shock is 
larger than the local density of the surrounding medium. As this first 
internal shock hits the bow-shock region and therefore continues to drive the 
propagation of the entire jet, this balance, i.e. the transition from an 
initially underdense to an overdense jet, leads to the substantial change in 
kinematics described above. The same scenario also holds for model iv', the 
velocity of the jet head is about 700 km s$^{-1}$.

\subsection{Internal jet structure}

As in the old model i with cooling, the propagation of all pulses again can be 
traced until their merger with the jet head. Each new pulse is instantaneously 
slowed down from 2000 km s$^{-1}$ to about 1200 km s$^{-1}$ within the first 
two days. The distance between the internal shocks created by the periodic 
velocity pulses stays constant. The very simple, periodic jet 
structure in the axial cuts of the Mach number, axial velocity, density, 
pressure, and temperature -- showing knots and the jet beam -- remains 
unaffected in this large scale simulation by the aforementioned transition.

A difference between model i' and the former model i with cooling is the 
radial extent of these knots (Fig. \ref{NV3.1den}). The whole width of the jet,
however, is again taken by the jet beam, i.e. by unshocked jet material. A 
cocoon of shocked jet matter is only a very small shell close to the contact 
discontinuity, and a backflow is again absent. Therefore the basic composition 
of the jet is not changed, only the dimensions. While in model i with cooling, 
the knots had a radius of 1 AU, their extent is now a few AU. The density 
level in the knots, however, is lowered by a factor of ten. Therefore the 
total mass of each knot could be constant during this spreading process, which 
could be triggered by a trough in the pressure profile at an intermediate jet 
length. There also, the temperature and the sound speed are reduced, leading 
to large Mach numbers.

The axial velocity component stays nearly constant inside the jet, but is 
slightly enhanced at radii of 5 AU compared to the jet axis. This leads to a
slow overtaking of the radial extensions of the jet pulses. The radial 
velocity component shows negative velocities, i.e. towards the axis, at larger 
distances near the jet axis. This seems to be another result of this 
overtaking, leading to an eddy-like structure near the jet head.

Another feature in the radial velocity is the linear, Hubble-like increase 
at lower distances from the jet axis. This is often seen in Planetary Nebulae 
\citep[e.g.][]{CLB01} and can be explained by a fast wind traveling 
into a self-similar $1/r^2$ density profile \citep[e.g.][]{KoM92}.
As the radial velocities are still of the order of a few hundred 
km s$^{-1}$, a radial radiative dense shell of shocked ambient medium is 
present.

\begin{figure*}
  \includegraphics[width=8cm]{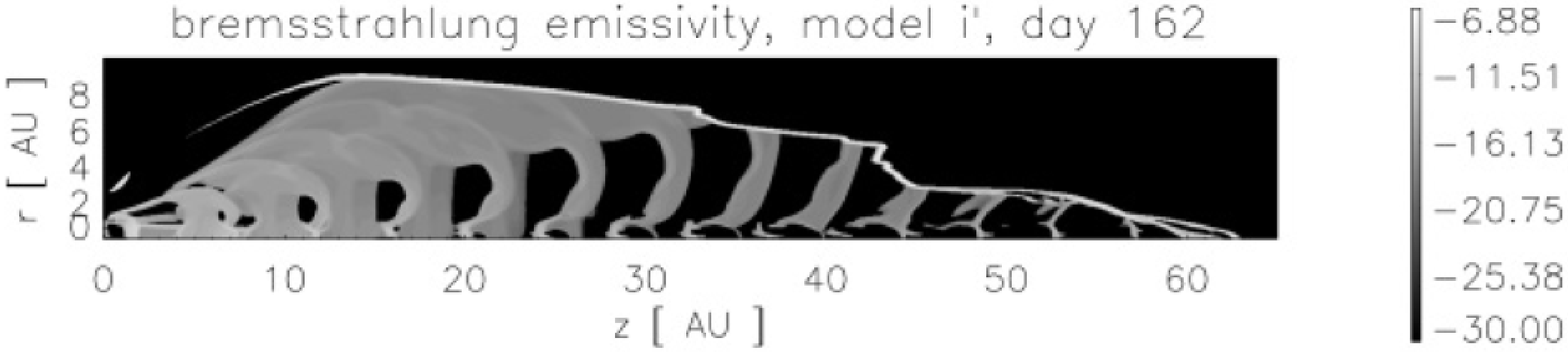}
  \includegraphics[width=8cm]{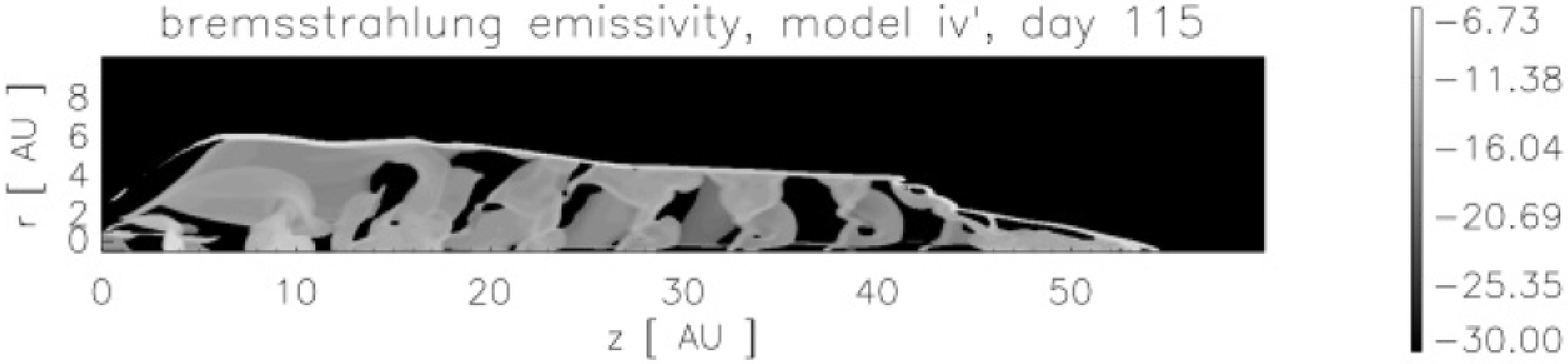}

  \includegraphics[width=8cm]{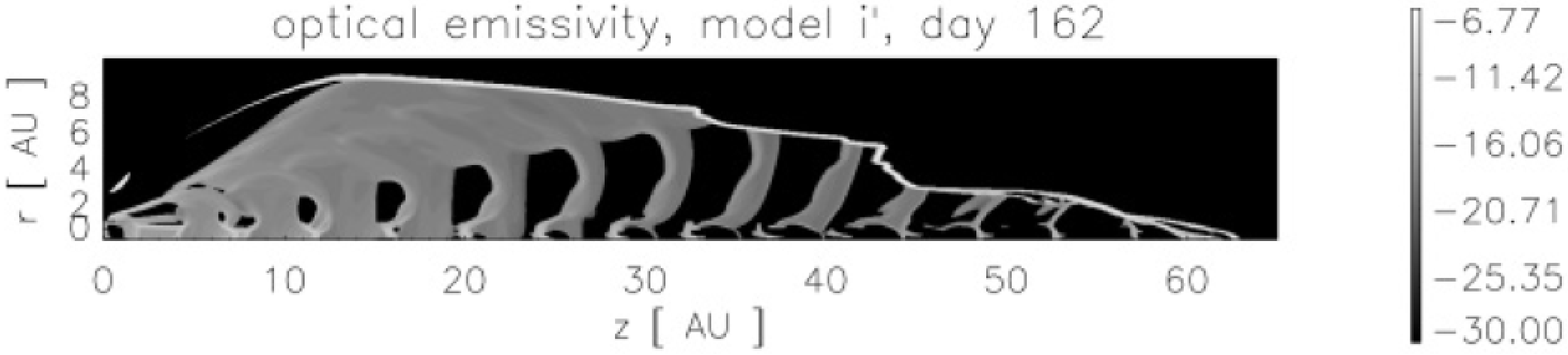}
  \includegraphics[width=8cm]{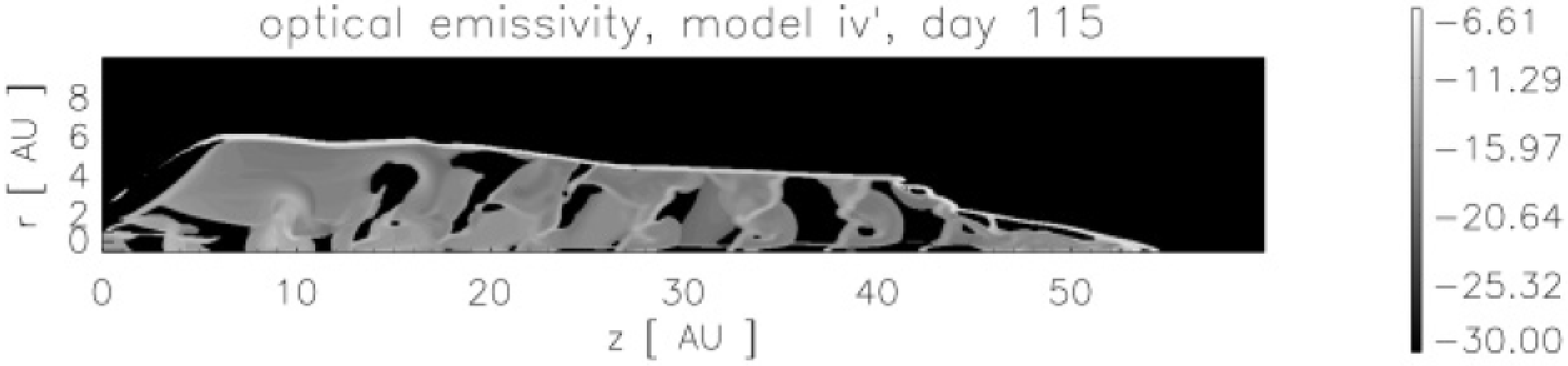}

  \includegraphics[width=8cm]{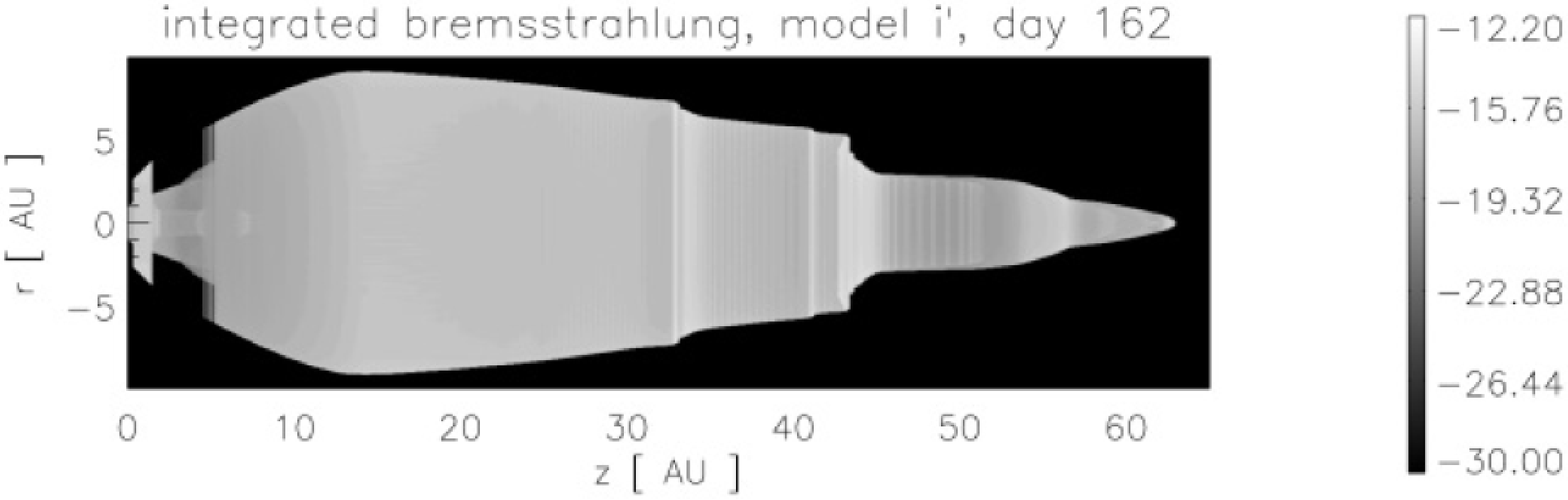}
  \includegraphics[width=8cm]{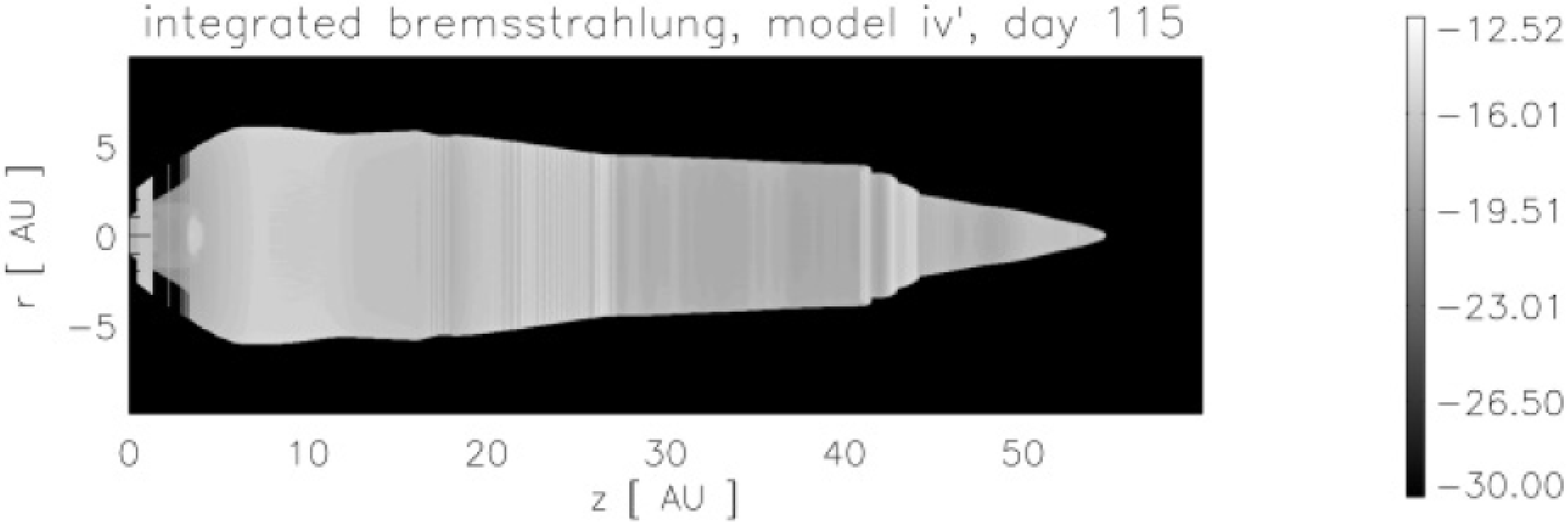}

  \includegraphics[width=8cm]{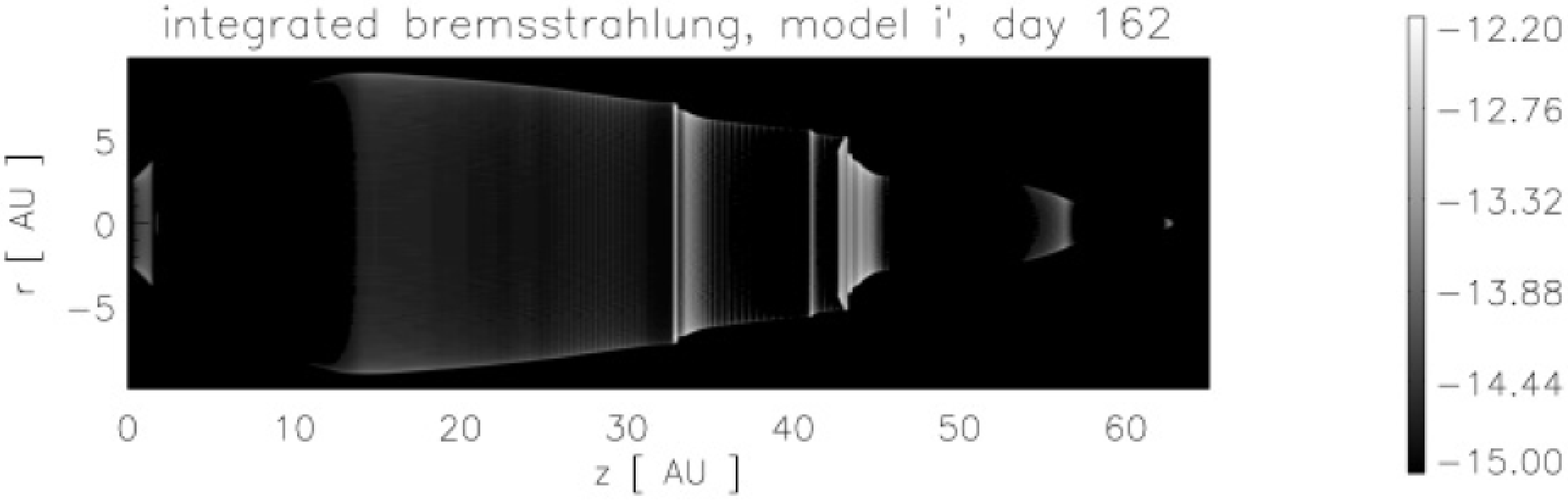}
  \includegraphics[width=8cm]{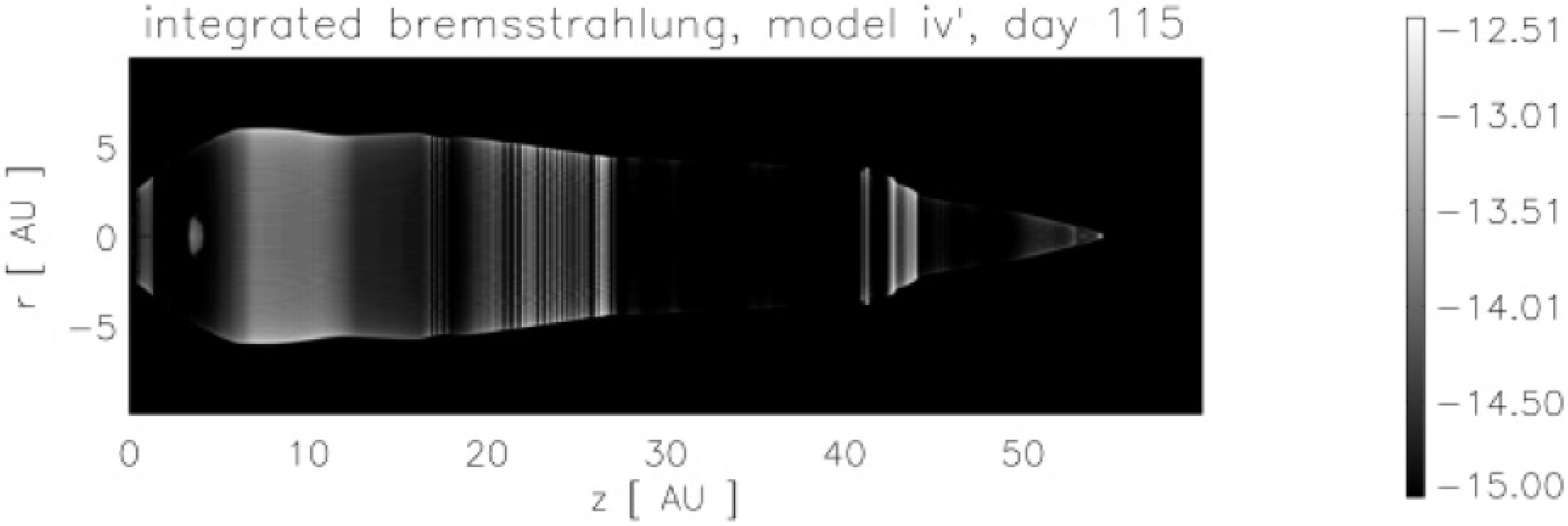}

  \includegraphics[width=8cm]{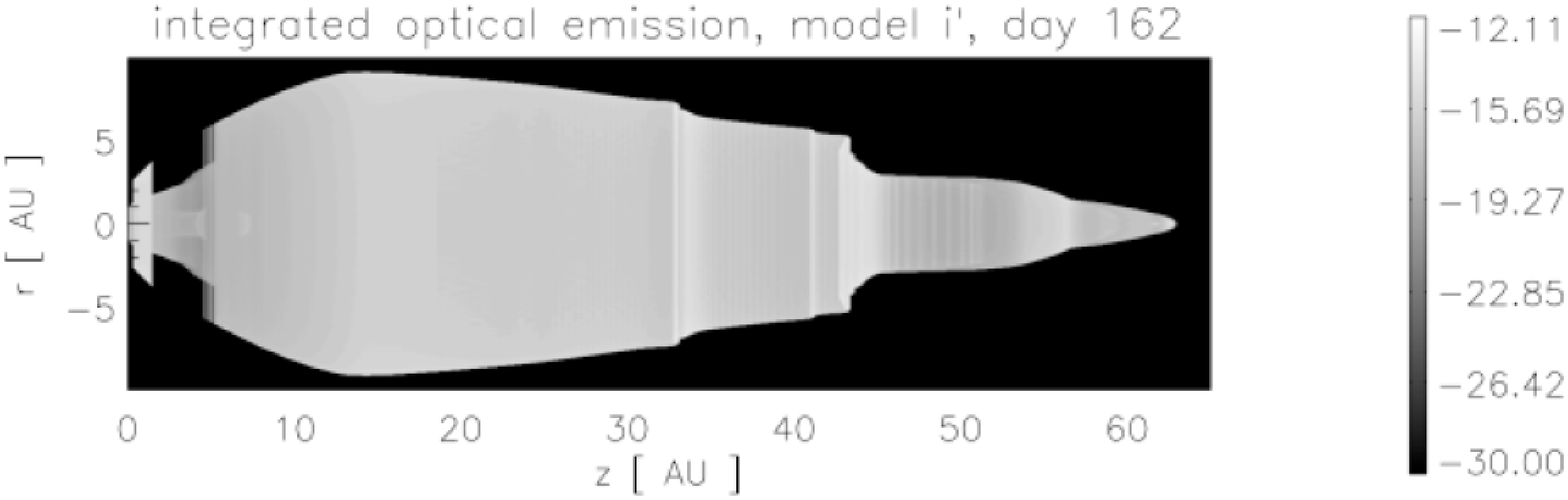}
  \includegraphics[width=8cm]{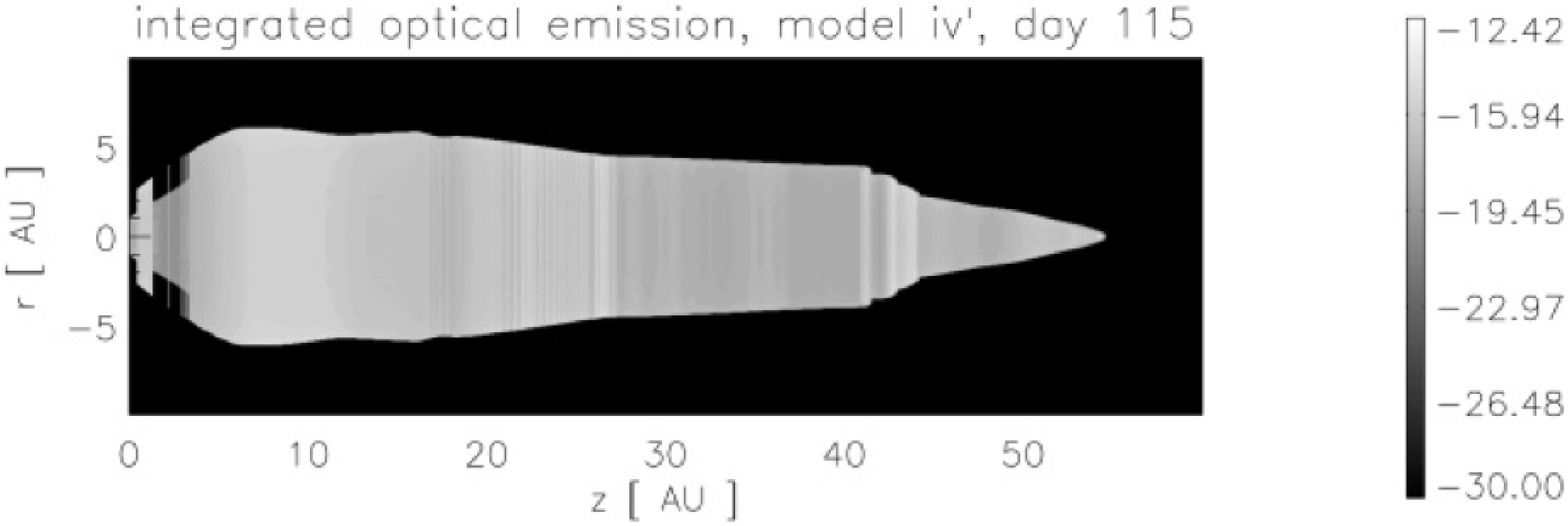}

  \includegraphics[width=8cm]{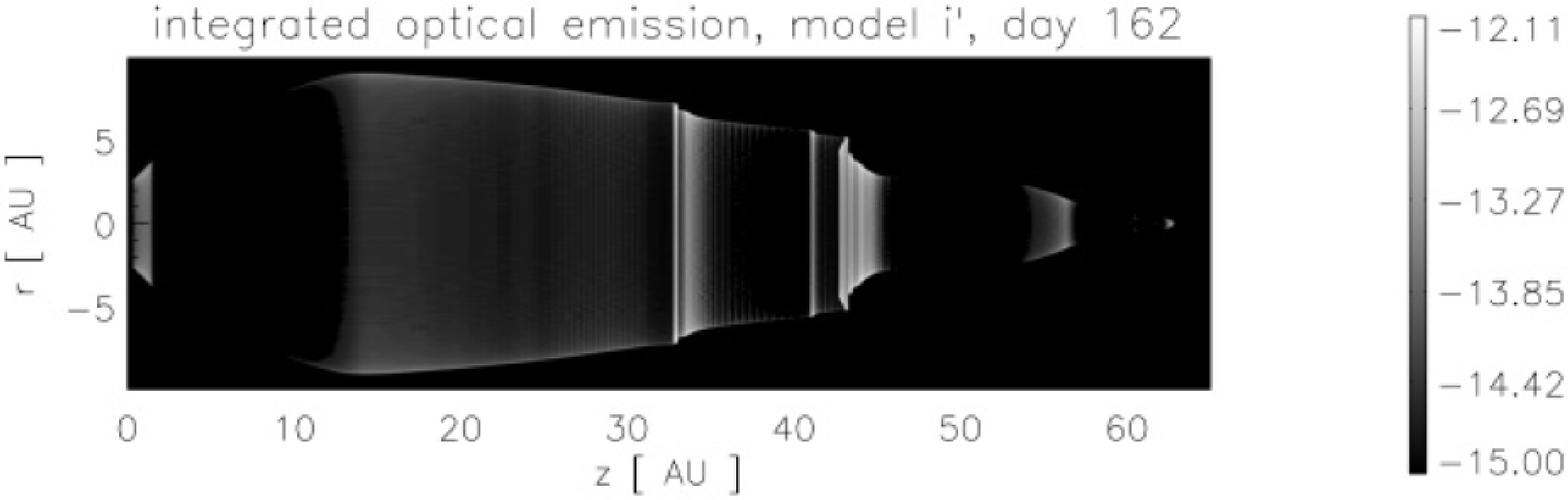}
  \includegraphics[width=8cm]{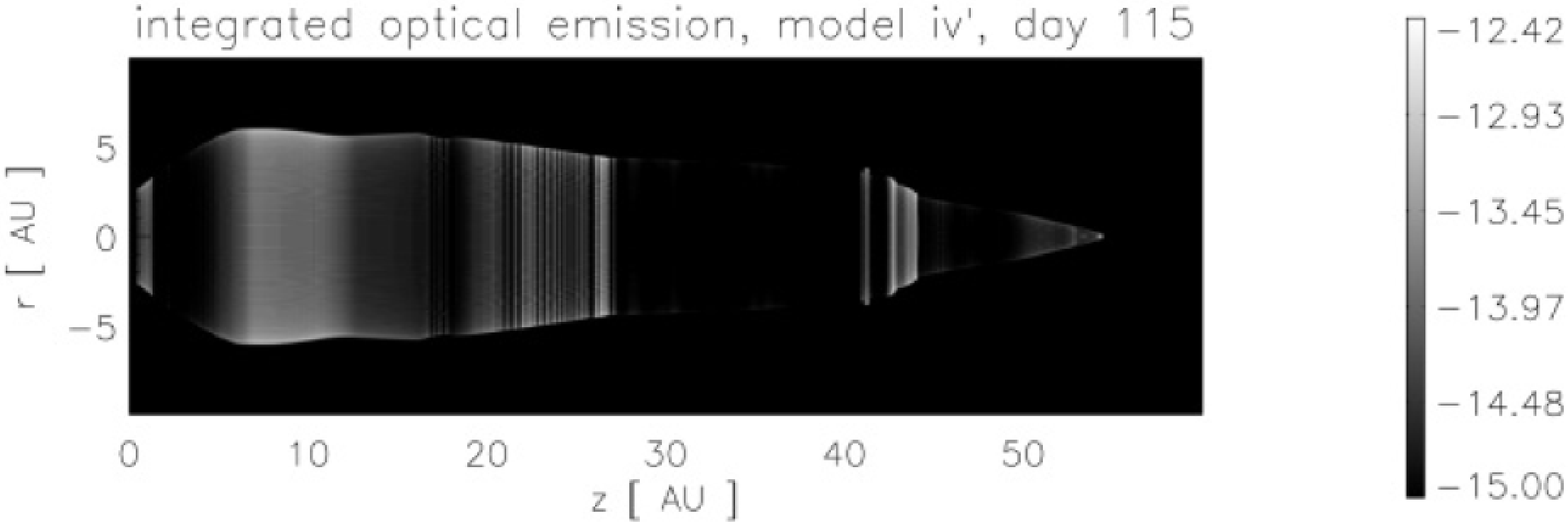}
  \caption{Radiative properties of model i' on day 162 (left) and model iv' on 
    day 115 (right): the emissivity of bremsstrahlung and optical 
    emission in erg s$^{-1}$ cm$^{-3}$, emission maps of bremsstrahlung and 
    optical emission in erg s$^{-1}$ cm$^{-2}$ with an assumed distance of 
    200 pc -- with and without a detection threshold of 
    $10^{-15}$ erg s$^{-1}$ cm$^{-2}$. The emission of the internal knots is 
    blended by the emission of the dense radiative shell of shocked ambient 
    medium}
  \label{NV3.1emiss}
\end{figure*}

In model iv', the internal knots are also extended in radial size, but are no
longer pronounced as in model i' along the full width of the jet. Instead 
of the clear crescent form in model i', the structure of the knots is now more 
complicated. Furthermore, they are drifting away from the jet axis while 
traveling and do not hit the jet head, but the lateral shock, leading to a 
sharp step. Therefore the pulses are not able to transfer the full momentum 
to the propagation of the jet head.

\section{Radiative properties of the jets} \label{sec_rad}

\subsection{Emission maps} \label{sec_emiss}

Using the density and pressure data in each grid cell of model i', we can 
calculate the temperature, the ionization fraction $X = n_{e} / n_{H}$, and 
then the total emissivity due to bremsstrahlung using the formula
\begin{equation}
j = 1.68 \cdot 10^{-27}\,T^{1/2}\,n_{e}\,n_{i}\, \rm{erg\,s}^{-1} 
\rm{cm}^{-3} 
\end{equation}
according to e.g. \citet{RyL}. We also calculate the optical emissivity of 
Balmer lines \citep{All} using
\begin{equation}
j = 4.16 \cdot 10^{-25} \, T_{4}^{-0.983}
\,10^{-0.0424/T_{4}}\,n_{e}\,n_{i}\, \rm{erg\,s}^{-1} \rm{cm}^{-3}
\end{equation}

\noindent
with $T_{4}$ the temperature in units of $10^4$ K. Both emissivities are 
plotted in the first two rows of Fig. \ref{NV3.1emiss}. 

These emissivities are then rotated and integrated along a line of sight that 
is perpendicular to the jet axis. Assuming a distance of 200 pc to the 
observer, emission maps are created (Fig. \ref{NV3.1emiss}, rows 3 and 5). The 
emission above a detection threshold of $10^{-15}$ erg s$^{-1}$ cm$^{-2}$ is 
also given.

In the emissivity plots, the knots with high density and lower temperature can 
be well detected because of their emission which is higher than that of the 
jet beam with its low density and higher temperature. The distinctness of 
these structures in the cocoon increases downstream towards the jet head and 
allows them to be detected more easily. After creating emission maps, 
however, the emission of the knots is blended by a strong emission of the 
dense radiative shell of shocked ambient medium.

The parts above the threshold show several distinct features that are created 
not by the knots, but by emission variations in this shell. As axisymmetry was 
assumed, these features are rings around the jet, observed as stripes at high 
inclinations.

\subsection{The absorption line profiles} \label{sec_abs}

In Paper I, the calculations of the jet absorption profiles have been 
presented in detail. Again, we start with a normalized continuum emission and 
a Gaussian emission line profile:
\begin{equation} \label{initial_line}
I_{0} = 1 + I_{\rm peak} \cdot \exp{\left( - v^2/\sigma^2 \right)} .
\end{equation}
The initial continuum and line emission (\ref{initial_line}) are taken as input
for the innermost grid cell of each path. We then calculate the absorption
within each grid cell $j$ along the line of sight according to:
\begin{eqnarray} \label{eq_abs_cs}
I_{j} &=& I_{j-1} \cdot e^{-\tau_\lambda}  \nonumber \\
 \tau_\lambda &=& \frac{\pi \, e^2}{m_{e} \, c} \,
 \lambda_{kl} \, \left( 1 - \frac{v_{j}}{c} \right) \,
 \frac{\rho_{j}\,\eta}{m_{H}} \, \Delta x_{j}\,f_{kl}  \nonumber \\
 &=& \mathcal{C}_{kl}\,\Delta x_{j}\,\mathcal{F}(v_{j},\rho_{j})\,,
\end{eqnarray}
\noindent
where $e$ is the electron charge, $m_{e}$ the electron mass, $c$ the speed of
light, $\lambda_{kl}$ the rest frame wavelength of the transition, $m_{H}$ the
proton mass, and $f_{kl}$ the oscillator strength. The parameters are constant
for a given atomic transition. We choose the \ion{Ca}{ii} K transition here 
with $\lambda_{kl}=3934$ \AA. $\Delta x_{j}$ is the length of the path
through grid cell $j$, which is a function of the inclination $i$.

The parameters depending on the different hydrodynamical models are the
velocity projected onto the line of sight $v_{j}$, the mass density $\rho_{j}$,
and $\eta$ the relative number density with respect to hydrogen
$\eta=n_k/n_{\rm H}$ of the absorbing atom in the lower level $k$ of the
investigated line transition. The velocity is binned for the calculation of
the absorption. The size of the bins $\Delta v$ can be interpreted as a
measure of the kinetic motion and turbulence in one grid cell. This absorption
dispersion helps to smooth the effects of the limited spatial resolution of
the numerical models. The absorption calculation through the jet region is 
then repeated for all possible light paths from the emission region to the 
observer. The arithmetic mean of the individual absorption line profiles from 
each path is then taken as the resulting spectrum.

In Paper I,  we described that the high-velocity components in the synthetic 
absorption line profiles are not as pronounced as in the observed ones. Now, 
our model iv' could be a solution to that problem, since both ways of 
increasing the absorption in the high velocity components -- higher pulse 
density and the cooling treatment -- are now present. The parameters are now 
chosen as in the best-fit model in Paper I: $I_{\rm peak}=6$, $\sigma=100$ km 
s$^{-1}$, $\lambda_{kl}=3934$ \AA, $f_{kl}=0.69$, $\eta=2 \cdot 10^{-6}$ for 
the \ion{Ca}{ii} K transition, $\Delta v =10$ km s$^{-1}$, $i=0^\circ$, and 
$r_{\rm em}=1$ AU. 

\begin{figure}
   \resizebox{\hsize}{!}{\includegraphics{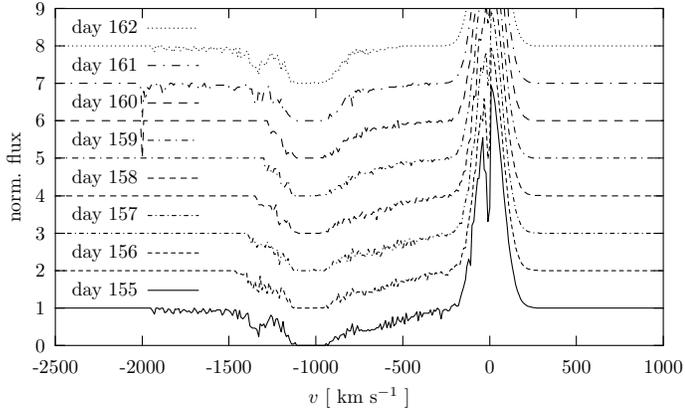}}
   \caption{Sequence of absorption line profiles for eight consecutive days 
     for model i'}
   \label{line_time_NV3.1}
\end{figure}
\begin{figure}
   \resizebox{\hsize}{!}{\includegraphics{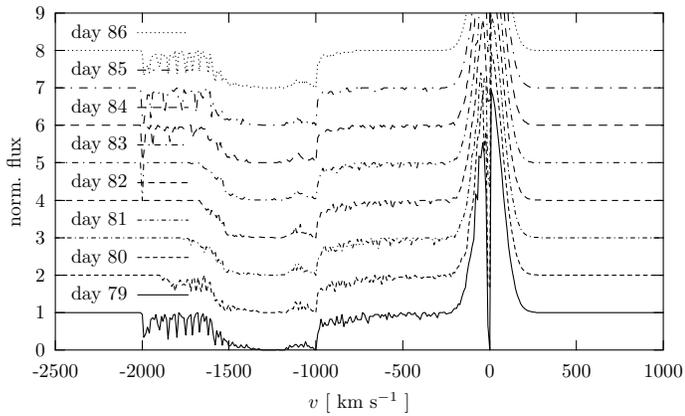}}
   \caption{Sequence of absorption line profiles for eight consecutive days of 
     model iv'}
   \label{line_time_NV3.2}
\end{figure}

In Figs. \ref{line_time_NV3.1}-\ref{line_time_NV3.2}, the sequences of eight
consecutive nights of the two large-scale models i' and iv' are shown. In the 
first case, the extension of model i' to longer propagation times and 
larger lengths create only slight differences compared to Fig. 26 (bottom) in 
Paper I. The position of the main absorption trough, the 
shape of its boundaries, the presence, persistence, and depth of the high 
velocity component are all very similar to those in the former model i with 
cooling. The absorption in the center of the emission line is 
created by the ambient medium in front of the jet.

The model iv' with increased jet pulse densities, however, creates clearly
visible differences. The stable absorption trough is shifted towards higher
velocities, from -1050 km s$^{-1}$ up to -1300 km s$^{-1}$, and its width is
increased from about 200 km s$^{-1}$ to 500 km s$^{-1}$. In the high velocity 
component, several deep absorptions are present in contrast to the few 
moderate ones in model i'. 

\section{Summary and discussion}

In this paper, we present results of two new hydrodynamical simulations, 
including radiative cooling. They were started due to a drawback in our former 
simulations presented in Paper I \citep{SCS}, which were stopped close to 
density balance of the jet and the ambient medium. As underdense and overdense 
jets behave in different ways and as the kinematics of the jets was a main 
result in Paper I, we expected new insights. 

Another point is that a few minor differences between the models and the 
observations were thought to be solved by a model with an increased jet density
during the pulses -- model iv' -- which was calculated only with purely 
hydrodynamical means in the former grid of simulations. Therefore, we 
performed two hydrodynamical simulations with an approximated cooling 
treatment beyond this density balance, one with the same parameters as model i 
in Paper I, which was presented there with and without cooling, and the second 
with higher gas densities in the jet pulses, as in model iv. 

Since in the new cooling treatment ionization fractions of collisional 
ionization equilibrium (CIE) were used and not self-consistent general 
non-equilibrium (NEQ) rate equations, we discussed the possible effects of 
underestimating the ionization fraction of hydrogen and validated this cooling 
treatment by comparing density and temperature values of the gas parcels in 
the jet, as well as by showing plots of the ionization fraction itself. We saw 
that the differences are only very small.

In Paper I, the jet in model i with cooling had a constant cross section over 
the simulated 74 days. This led to an accelerated motion with high velocities 
of 730 km s$^{-1}$, while the velocity of the adiabatic jet was only about 200 
km s$^{-1}$. The extended model i' in this paper now shows that after the 
first 70 days the transition from an underdense to an overdense jet lets the 
cross section inflate and therefore compensate for the density profile of the 
external medium. The acceleration is stopped and the jet then has a constant 
velocity of about 740 km s$^{-1}$. Thus, the transition results in a 
completely different motion. The conclusion that the high observed velocities 
in CH Cygni, R Aquarii, and MWC 560 favor the models with cooling is unchanged 
by the transition.

The radial inflation mentioned above also changes the internal structure of the
jet. The width of the internal knots is also increased, as the basic 
composition of the jets is not affected. The cocoon of shocked jet matter is 
again confined to a small shell close to the contact discontinuity and the 
whole jet lobe is filled with unshocked jet material. In model i', these knots 
are very pronounced along their width, while their structure is not intact in 
model iv'.

These knots are the locations of enhanced bremsstrahlung and optical 
emissivity, which is therefore again more pronounced in model i'. These knots 
could be identified with the observed parallel features in R Aquarii 
\citep{PaH}. After rotating the emissivity plots and integrating them, however,
other features become more prominent. The internal knots are blended by the 
emission of the dense radiative shell of shocked ambient medium. As they are 
spatially variable, parallel rings of enhanced emission are created by the 
rotation and also look similar to the observations, if one keeps in mind that 
axisymmetry was assumed in our model. Without that, the shapes of the features 
on the shell should be different.

The observed length of the jet in R Aquarii is several hundred AU, thus much 
larger than in our simulations. At larger distances from the source, however, 
the density of the ambient medium drops by another order of magnitude, making 
it likely that the relative contribution of the emission of the knots will 
increase. 

Concerning the absorption line profiles, no large differences between the 
former model i with cooling and the recent model i' are visible. This means 
that the influence of the age of the jet on the profiles should only be 
marginal. In model iv', however, differences arise. The stable absorption 
trough is shifted toward higher velocities, from -1050 km s$^{-1}$ up to 
-1300 km s$^{-1}$, and its width is increased from about 200 km s$^{-1}$ to 
500 km s$^{-1}$. In the high velocity component, several deep absorptions are 
present in contrast to the few moderate ones in model i'. 

Both results highly increase the ability of the model to reproduce the 
observed absorption line profiles \citep[see Fig.1 in Paper I or ][]{SKC}. We 
can say here that the real parameters in MWC 560 are closer to model iv' than 
to model i'.

As the pulse structure was not changed, the line widths of the high velocity
components are again quite narrow. This could perhaps be improved by using a
sinusoidal or Gaussian pulse profile, instead of the rectangular steps in
velocity and density, and also a spatial velocity and density profile, instead 
of the constant values within the jet nozzle.

\begin{acknowledgements}
The author wants to thank the High Performance Computing Center Stuttgart for 
allowing him to perform the expensive computations, and H.M. Schmid and M.
Camenzind for fruitful discussions. We acknowledge the constructive comments 
and suggestions by the referee. 
\end{acknowledgements}

\end{document}